\documentclass[fleqn,usenatbib,usedcolumn]{mnras}
\usepackage[british]{babel}
\usepackage{graphicx}
\usepackage{color,soul}
\usepackage{mathtools}
\usepackage{newtxtext}
\usepackage[slantedGreek]{newtxmath}
\usepackage[T1]{fontenc}
\usepackage{bm}
\usepackage{amsmath}
\usepackage{hyperref}
\usepackage{booktabs}
\usepackage[inline]{enumitem}
\usepackage{relsize}
\usepackage{todonotes}

\addto\extrasbritish{%
}

\let\mr\mathrm
\graphicspath{{Figures/}}

\newcommand{\sTheta}{\bm{\upTheta}} 
\newcommand{\sdelta}{\deltaup}      

\begin{document}
\title[Testing for calibration systematics in the EDGES low-band data using Bayesian model selection]{Testing for calibration systematics in the EDGES low-band data using Bayesian model selection}
\author[Sims et al.]{Peter H. Sims$^1$\thanks{E-mail: peter\_sims1@brown.edu}, Jonathan C. Pober$^1$ \\
$^1$Department of Physics, Brown University, Providence, RI 02912, USA \\
}

\maketitle
\label{firstpage}
\begin{abstract}
Cosmic Dawn, when the first stars and proto-galaxies began to form, is commonly expected to be accompanied by an absorption signature at radio frequencies. This feature arises as Lyman-$\alpha$ photons emitted by these first luminous objects couple the 21 cm excitation temperature of intergalactic hydrogen gas to its kinetic temperature, driving it into absorption relative to the CMB. The detailed properties of this absorption profile encode powerful information about the physics of Cosmic Dawn. Recently, Bowman et al. analysed data from the EDGES low-band radio antenna and found an unexpectedly deep absorption profile centred at 78 MHz, which could be a detection of this signature. Their specific analysis fit their measurements using a polynomial foreground model, a flattened Gaussian absorption profile and a white noise model; we argue that a more accurate model, that includes a detailed noise model and accounting for the effects of plausible calibration errors, is essential for describing the EDGES data set. We perform a Bayesian evidence-based comparison of models of the EDGES low-band data set and find that those incorporating these additional components are decisively preferred. The subset of the best fitting models of the data that include a global signal favour an amplitude consistent with standard cosmological assumptions ($A<209~\mathrm{mK}$). However, there is not strong evidence to favour models of the data including a global 21 cm signal over those without one. Ultimately, we find that the derivation of robust constraints on astrophysics from the data is limited by the presence of systematics.
\end{abstract}


\section{Introduction}
\label{Introduction}

The redshifted 21 cm hyperfine line emission from the neutral hydrogen that pervaded the intergalactic medium (IGM) during its transition from a cold and predominantly neutral state at the end of the cosmic dark ages (DA; $z\sim20$) to a hot, ionized plasma by the end of the Epoch of Reionization (EoR; $z\sim6$) is a powerful probe of the intervening period (see e.g. \citealt{2006PhR...433..181F, 2007RPPh...70..627B, 2010ARA&A..48..127M, 2012RPPh...75h6901P}). The intensity of the redshifted 21 cm emission can be measured as a differential brightness temperature between the hydrogen spin temperature\footnote{The spin temperature quantifies the relative number densities, $n_{1}$ and $n_{0}$, of hydrogen in the spin triplet ($F_1$) and spin singlet ($F_0$) hyperfine levels of the electronic ground state via the Boltzmann equation:
$\frac{n_1}{n_0}=3e^{-T_\star/T_\mathrm{S}}$,
where the factor 3 is the ratio of the statistical weights (i.e. energy degeneracy) of the triplet and singlet states and $T_\star = 0.068~\mathrm{K}$ is the excitation temperature of the spin-flip transition.} and the brightness temperature of the radio background (e.g. \citealt{2006PhR...433..181F}),
\begin{equation}\label{Eq:DifferentialBrightness} 
\sdelta T_\mathrm{b} \approx 9~\mathrm{mK}~x_{\ion{H}{i}}(1+\delta_\mathrm{b}) (1+z)^{1/2} \left[1-\frac{T_\mathrm{R}(z)}{T_\mathrm{S}}\right]\left[\frac{H(z)/(1+z)}{\mathrm{d}v_\parallel / \mathrm{d}r_\parallel}\right] \ ,
\end{equation}
where $\delta_\mathrm{b}$ is the fractional baryon overdensity, $x_{\ion{H}{i}}$ is the hydrogen neutral fraction, $H(z)$ is the Hubble parameter, $\mathrm{d}v_\parallel / \mathrm{d}r_\parallel$ is the gradient of the proper velocity along the line of sight, $T_\mathrm{R}$ is the background radiation temperature and the hydrogen spin temperature, $T_\mathrm{S}$, is given by,
\begin{equation}
T_S = \frac{1 + x_c + x_\alpha}{T_\mathrm{R}^{-1} + x_c T_K^{-1} + x_\alpha T_c^{-1}} \ .
\label{Eq:SpinTemp}
\end{equation}
Here, $x_c$ and $x_\alpha$ are coupling coefficients for collisions and UV scattering and $T_K$ and $T_c$ are the hydrogen kinetic temperature and the effective color temperature of the UV radiation field, respectively. $\sdelta T_\mathrm{b}$ thus provides a direct probe of the ionisation, density and temperature state of the IGM during this period.

In fiducial models (e.g. \citealt{2008PhRvD..78j3511P}), the background radiation temperature is dominated by the cosmic microwave background (CMB): $T_\mathrm{R}\simeq T_{\gamma}$, with $T_\mathrm{R}$ and $T_{\gamma}$ the background and CMB radiation temperatures, respectively. The kinetic temperature of the hydrogen in the IGM, $T_\mathrm{K}$, decouples from the CMB radiation temperature at redshift $z\approx150$. It subsequently cools via adiabatic expansion at a rate of $(1+z)^{2}$ relative to $T_{\gamma}$, which drops at the slower rate of $(1+z)$. Initially, there is a sufficient density of free electrons in the IGM to couple $T_{S}$ to $T_{K}$, leading to an early absorption trough; however, continuing expansion reduces the strength of this coupling and, by $z\sim70$, radiative coupling of $T_{S}$ to $T_{\gamma}$ is expected to dominate, resulting in $\sdelta T_\mathrm{b}\sim0$.

Cosmic Dawn (CD), when the first stars and proto-galaxies began to form, produces the next key milestone in the evolution of the state of the IGM. The first sources emit photons at the Ly$\alpha$ resonance of hydrogen. As Ly$\alpha$ photons permeate the hydrogen gas in the IGM, their absorption and spontaneous re-emission recouples $T_{S}$ to $T_{K}$, via the Wouthuysen--Field effect\footnote{More directly, the Wouthuysen--Field effect couples $T_{S}$ to $T_{c}$; however, the high optical depth of the Ly$\alpha$ line means the radiation colour temperature and kinetic temperature of the gas are rapidly equilibrated and, thus, the two temperatures are approximately equal in practice (e.g. \citealt{2006PhR...433..181F}).} (\citealt{1952AJ.....57R..31W, 1958PIRE...46..240F}). The corresponding cooling of the spin temperature below the CMB temperature produces an absorption signature in the observed 21 cm brightness temperature. At the same time, Lyman continuum photons emitted by the first sources begin the process of reionizing the IGM and X-ray photons, for example, from early black hole binary systems (e.g. \citealt{2014Natur.506..197F}), raising the kinetic temperature of the gas, reducing the strength of the absorption signature and, subsequently, generating a signal in emission, if, prior to the completion of reionization, the temperature of the gas is raised in excess of the CMB temperature.

As a result, the amplitudes of the signals in both absorption and emission are a sensitive function of the time dependent strengths of the Ly$\alpha$, Lyman continuum and X-ray radiation fields during this period. The strengths of these fields are poorly constrained by current data; however, the maximum depth of the absorption trough in fiducial models can be derived, assuming adiabatic cooling of $T_{K}$ between $z\approx150$ and CD, no reheating of the gas prior to the Wouthuysen--Field coupling of $T_{S}$ to $T_{K}$ and $T_\mathrm{R} \approx T_{\gamma}$ at CD. In this case, an absorption trough with a depth of $209~\mathrm{mK}$ is expected (\citealt{2018Natur.555...71B}). 

Two distinct varieties of experiments aimed at measuring the evolution of the sky-averaged `global' redshifted 21 cm signal and fluctuations in its intensity as a function of spatial scale, respectively, are underway. Instruments in the first category include: EDGES (\citealt{2008ApJ...676....1B}), SCI-H I (\citealt{2014ApJ...782L...9V}), BIGHORNS (\citealt{2015PASA...32....4S}), LEDA (e.g. \citealt{2016MNRAS.461.2847B}) and SARAS (\citealt{2018ApJ...858...54S}). While those currently in operation in the second category include: the GMRT (\citealt{2013MNRAS.433..639P}), LOFAR (\citealt{2013A&A...556A...2V}), the MWA (\citealt{2013PASA...30....7T}), PAPER (\citealt{2010AJ....139.1468P}) and HERA (\citealt{2017PASP..129d5001D}), while the SKA (\citealt{2013ExA....36..235M}) is set to commence in the near future.

The global signal, targeted principally by single-dipole experiments, traces the sky-averaged ionization and thermal history of the hydrogen IGM and constrains the timing of CD and the EoR. In 2018, analysing data in the 50--100 MHz range, the EDGES experiment reported the first detection of the global 21 cm signal (\citealt{2018Natur.555...67B}; hereafter B18). They recover a best fitting flat bottomed absorption trough centred at $78 \pm 1~\mathrm{MHz}$, with a width of $19^{+4}_{-2}~\mathrm{MHz}$ and with a depth of $500^{+500}_{-200}~\mathrm{mK}$, where the uncertanties correspond to 99 per cent confidence intervals, accounting for both thermal and systematic errors. The depth of the absorption trough recovered with this model is more than double the maximum depth predicted by fiducial models at a $\sim3.8~\sigma$ significance (B18). The flat bottomed shape of the absorption trough is also unexpected and suggests that $\mathrm{Ly}\alpha$ photons flooded most of the universe within a small fraction of a Hubble time, followed by an extended period with little to no heating of the IGM and, finally, the gas temperature quickly rose and diminished the absorption signal (\citealt{2018ApJ...864L..15K}).

If it is assumed that the absorption parameters derived with this model are accurate, its large depth implies an increased differential brightness between the hydrogen spin temperature and the radio background, relative to the fiducial model described above. From \autoref{Eq:DifferentialBrightness}, $\sdelta T_\mathrm{b} \propto (1-T_\mathrm{R}(z)/T_\mathrm{S})$ and thus, relative to the standard cosmological model, where $T_\mathrm{R} = T_{\gamma}$, an enhanced absorption depth will result if either \begin{enumerate*}\item the radio background temperature is larger than expected, which would occur if there was an unaccounted for radio background in excess of the CMB ($T_\mathrm{R} > T_{\gamma}$; B18; \citealt{2018ApJ...858L..17F}), or \item if $T_{K}$, and correspondingly $T_{S}$, is cooled to below the level expected from adiabatic cooling alone (B18; \citealt{2018Natur.555...71B}).
\end{enumerate*}

Raising the radio background temperature above the CMB radiation temperature would require the presence of a bright, unaccounted for source of radio emission at CD. Potential candidates for such sources include: \begin{itemize}\item synchrotron emission\footnote{However, see \citet{2018MNRAS.481L...6S} for the impact of a reduced radiation cooling time (given by the shorter of the synchrotron and inverse-Compton cooling times) of the relevant relativistic electrons, at $z\sim17$ compared to in the local Universe.} from accretion onto a population of growing black holes (\citealt{2018ApJ...868...63E}), \item enhanced radio emission from the first Galaxies\footnote{The $500~\mathrm{mK}$ absorption depth necessitates a three order of magnitude increase in the efficiency of photon production in the $\sim1$--$2~\mathrm{GHz}$ spectral range, relative to star forming galaxies today.} (\citealt{2019MNRAS.483.1980M}), \item soft photon emission from the decay of axionic dark matter decay in mini-clusters (\citealt{2018PhLB..785..159F}), \item the decay of dark matter to dark photons, followed by resonant oscillations of dark photons into regular photons in the Rayleigh-Jeans tail of the CMB (\citealt{2018PhRvL.121c1103P}).                                                                                                                                                                                                               \end{itemize}
Regarding these hypotheses, it is also notable that a radio background excess has previously been reported by the ARCADE2 experiment (\citealt{2011ApJ...734....5F}) and, more recently, by the LWA1 (\citealt{2018ApJ...858L...9D}). If only a small fraction of the excess emission reported by these experiments originated at CD, it would be sufficient to account for the EDGES absorption depth\footnote{However, it has also been argued that the `excess' seen by these experiments, where the excess is defined as the difference between the measured emission and the (model-dependent) sky temperature expected from the total emission from the Galaxy, extragalactic sources and the CMB, may be artificial and, in fact, may be removed by better modelling of the contribution of Galactic emission to the radio background (\citealt{2013ApJ...776...42S}).} (e.g. \citealt{2018ApJ...858L..17F}).
 
Alternatively, mechanisms for cooling $T_{K}$ below the level expected from adiabatic cooling alone have been proposed, including: \begin{itemize} \item supplementing the adiabatic cooling of $T_{K}$ via a non-standard Coulomb-like interaction (with a scattering cross-section $\sigma \propto v^{-4}$, with $v$ the Baryon - dark matter relative velocity) between baryons and cold dark matter particles with kinetic temperatures below the Hydrogen kinetic temperature at CD (\citealt{2018Natur.555...71B}). However, this scenario is strongly constrained by observations (e.g. B18; \citealt{2018PhRvL.121a1102B}, \citealt{2018Natur.557..684M}). \item decoupling of $T_{K}$ from $T_{\gamma}$ before $z\sim150$, giving the gas additional time to cool adiabatically before the first luminous sources form. `Early dark energy'(EDE; an additional component with equation of state $w=-1$), that contributes to the cosmological energy density at early times, before decaying rapidly (\citealt{2018JCAP...08..037H}), is a mechanism that has been investigated as a means to facilitate such early decoupling. However, it is found to be strongly ruled out by observations of the CMB temperature power spectrum and, moreover, EDE models needed to explain the EDGES signal exacerbate the current tension in low- and high-redshift measurements of the Hubble constant (\citealt{2018JCAP...08..037H}).
\end{itemize}

Despite the exciting potential implications for dark matter, dark energy models and galaxy formation models that open up if the absorption parameters derived by B18 are correct, the extreme challenge posed by strong foreground emission, which necessitates exquisite calibration of the data and control over instrumental systematics, means that assessing the robustness of the modelling used by B18 to derive the global 21 cm signal is of significant importance.

Recently, \citet{2018Natur.564E..32H} (hereafter, H18) re-examined the fitting process used by B18 and concluded that for the range of models they used to analyse the EDGES low-band data, including the model used by B18, there is severe degeneracy between signal and foreground modelling that allows a wide range of `signals' to be consistent with the data. This calls into question the interpretation of the B18 signal as an unambiguous detection of the cosmological 21-cm absorption signature. Furthermore, H18 highlight the possibility of an uncalibrated sinusoidal systematic in the data.

Further evidence for the model dependence of the global 21 cm signal can be seen in the results of recent work by \citet{2019arXiv190304540S} (hereafter S19), who, building on the analysis of H18, fit for a maximally smooth polynomial foreground model, constant amplitude sinusoidal systematic model, white noise and a Gaussian parametrisation of the global 21 cm signal, and find that, in this case, a low-RMS residual fit to the data can be derived with an absorption depth not exceeding the maximum depth obtainable in fiducial cosmological models ($A \sim 100~\mathrm{mK}$). 

However, in this work, we will argue that the use of a white noise model for the EDGES low-band data is an oversimplification and that a more general model for systematic effects in the data is required. We, therefore, extend the analyses of B18, H18 and S19 to use physically motivated models for the noise on the data, the global 21 cm signal and plausible miscalibration systematics. Using these models and comparing them to those considered thus far in the literature, we perform a more comprehensive investigation of the feasible model space of the EDGES data and use a Bayesian evidence based comparison of the models to identify those that best describe the data.

The remainder of this paper will be organised as follows. In \autoref{Models}, we begin by examining the chromaticity-correction component of the calibration of the publicly available EDGES low-band data and the systematic structures it can impart to the data, if the beam or sky models used in the correction are imperfect. The possibility of such structure in the data motivates us to compare models with and without components designed to describe systematic effects. We follow this by describing the full complement of deterministic and stochastic models for the foregrounds, systematic effects, global signal and noise on the data that we select from to construct the models to be compared in the analysis, and their implementation in a Bayesian model selection framework. In \autoref{ModelEvidences}, we present the model evidence results obtained from our analysis. We analyse the physical parameters of highest evidence models in more detail in \autoref{AnalysisOfPreferredModels} and present our conclusions in \autoref{Conclusions}.

\section{Modelling the EDGES low-band data}
\label{Models}

In this section, we start by reviewing the chromaticity correction element of the calibration of the EDGES low-band data in \autoref{EDGESData}. Since this correction can potentially result in systematic effects in the data, we compare models with and without components designed to describe these effects, to determine which models best describe the data. A Bayesian evidence based approach provides a statistically robust framework for performing such a comparison of models for the EDGES low-band data. Such an analysis is based on the principles of Bayesian inference, which we discuss in \autoref{BayesianInference}. Finally, in \autoref{EDGESDataModel} we include a description of the full set of deterministic and stochastic components of the models for the EDGES low-band data that we employ in our analysis.

\subsection{EDGES low-band data}
\label{EDGESData}

In this work, we use the publicly-released EDGES low-band spectrum available at: \href{http://loco.lab.asu.edu/edges/edges-data-release}{http://loco.lab.asu.edu/edges/edges-data-release}; this is the same data set used to derive the global 21 cm signal quoted by B18.
To derive their result, B18 calibrate the raw data from the EDGES low-band instrument and jointly fit a model to the calibrated data, consisting of a physically motivated five term polynomial foreground model and a flattened Gaussian parametrisation for the 21 cm signal. They approximate the noise on the data as spectrally flat. 

The data collection with the EDGES low-band experiment and its calibration are described in detail in B18. Here, we focus only on the beam chromaticity correction element of the data calibration, since imperfections in the beam model or sky model used in this step can both impart new and leave residual oscillatory spectral structure in the spectrum. This can have important implications for the data analysis and partially motivates our modelling choices in \autoref{EDGESDataModel}.

\subsubsection{EDGES beam chromaticity correction}
\label{EDGESBeamChromaticityCorrection}
The `adjustment for beam chromaticity' element of the EDGES data calibration is described in detail in \citet{2017MNRAS.464.4995M, 2019MNRAS.483.4411M}. The purpose of this procedure is to divide out the effect of beam chromaticity in the measured spectra using electromagnetic simulations of the beam, in combination with an approximate sky model. To aid discussion, we include a summary here for reference.

For a given LST value, a beam correction factor is obtained using an electromagnetic simulation of the beam and the 408~MHz Haslam all-sky map. This factor is calculated as, 
\begin{equation} 
\label{Eq:beam_correction}
B_{\text{factor}}(\nu) = \dfrac{\int_{\Omega}T_\text{sky-model}(\nu_{_{75}},\Omega) * B(\nu,\Omega) \mathrm{d}\Omega}{\int_{\Omega}T_\text{sky-model}(\nu_{_{75}},\Omega) * B(\nu_{_{75}},\Omega) \mathrm{d}\Omega } \ .
\end{equation}
Here,
\begin{equation} 
\label{Eq:sky_correction}
T_\text{sky-model}(\nu_{_{75}},\Omega) = \left [ T_\text{Haslam}(\Omega) - T_\text{CMB}\right ] \left ( \frac{75}{408} \right )^{-2.5}  + T_\text{CMB},
\end{equation}
$B(\nu_{_{75}}, \Omega)$ is the beam directivity at 75~MHz for a given pointing and orientation, $\nu$ is frequency, $\Omega$ are the spatial coordinates above the horizon, $T_\text{Haslam}(\Omega)$ is the 408~MHz Haslam all-sky map, and the CMB temperature, $T_\text{CMB}$, is $2.725~\mathrm{K}$. Unique beam correction factors are generated for 72 LST time intervals. The measured sky spectra are then corrected by dividing by the beam correction factor.

However, accurate removal of the chromatic structure in the beam via this procedure requires that: \begin{enumerate}\item the spatial structure of the foreground in the $50 - 100~\mathrm{MHz}$ range can be adequately approximated by the spatial structure at $75~\mathrm{MHz}$, \item the spatial structure at $75~\mathrm{MHz}$ is the same as that at $408~\mathrm{MHz}$ and \item the antenna's beam directivity is known to high precision in the $50 - 100~\mathrm{MHz}$ range.
\end{enumerate}

An in-depth investigation of the effect of realistic deviations from these assumptions would be necessary to determine how accurately each must be met for the beam chromaticity correction to reduce chromatic beam effects to below the noise level in the EDGES low-band data. Ideally, the best fitting model for the sky and beam would be jointly sampled over, within their respective estimated uncertainties, in order for accurate uncertainties on the global signal to be derived, but this requires access to the unaveraged data. In the absence of such an investigation, or joint analysis, it seems reasonable to assume a priori that existing chromatic structure in the data may not be perfectly removed and, indeed, that additional chromatic structure could be imparted by such a correction. In such a case, models for such systematic structure must be included in the data analysis, if we hope to recover robust estimates of the global signal that are free from bias.

\subsubsection{Chromatic structure imparted in the case of an imperfect chromaticity correction}
\label{EDGESBeamChromaticityCorrectionErrors}

The frequency dependence of the FEKO electromagnetic beam model of the linear gain of the EDGES beam at multiple beam angles is illustrated in B18's Extended Data Figure 4 b, and approximately sinusoidal undulations with frequency after a five-term polynomial has been subtracted from each of the gain curves are shown in Extended Data Figure 4 c. The amplitude, phase and period of the undulations are a function of the position in the beam. The amplitudes are of order 1\% of the gain amplitude, and the periods vary between approximately 20 MHz and 12 MHz at the locations in the beam shown. The amplitude and period both decrease with increasing angle from zenith.

For a sky temperature of the order $10^{3}~\mathrm{K}$, such undulations correspond to an amplitude feature of order $10~\mathrm{K}$ in the spectrum. Thus, the presence of such a feature must be correctly modelled to better than approximately one part in $10^{3}$ in amplitude across the beam for uncertainty in the model not to have a statistically significant impact on the data relative to the order $10~\mathrm{mK}$ noise level expected in the publicly available EDGES data set (see \autoref{CovarianceMatrixModel} for a discussion of the expected noise level).

We note that imperfections in the beam model\footnote{Equivalent systematics can similarly derive from imperfections in the sky model used in the chromaticity correction, or from errors in both the sky and beam model, but, here, for illustration, we principally focus on errors due to the mismodelling of the beam alone.} used in the beam chromaticity correction element of the data calibration, in excess of the level described above, can produce unwanted spectral structures in the data. The calibration systematics imparted in such a case will fall on a spectrum between two limiting cases:
\begin{enumerate}
 \item \textit{Beam model error localised in angle from zenith - } assuming a comparatively accurate model for the beam at the majority of angles from zenith, a frequency independent error in the amplitude of the undulating component of the electromagnetic beam model at a localised angle from zenith will result in a single sky temperature scaled sinusoid systematic feature in the data - i.e., a power law damped sinusoid with an expected power law index that is a function of the foreground spectra in the data and the shape of the gain curve. 
 \item \textit{Beam model error dispersed over angle from zenith - } if, instead, there are low level errors in the beam model spread over a wider range of beam angles, variation in the phase and period of the undulating gain feature with angle from zenith would result in multiple sky temperature scaled sinusoidal systematic features in the data, that, depending on their period, could be modelled, in principle, with multiple power law damped sinusoids. However, if a significant, and unknown, number of such features are present in the data, a generalised model for intermediate scale fluctuations, such as a log polynomial model with sufficient degrees of freedom to model structure on the scales of the intrinsic damped sinusoidal systematics and their beat frequencies, can provide a more compact parametrisation of the systematic.
\end{enumerate} 

\begin{figure}
	\centerline{
	\includegraphics[width=0.5\textwidth]{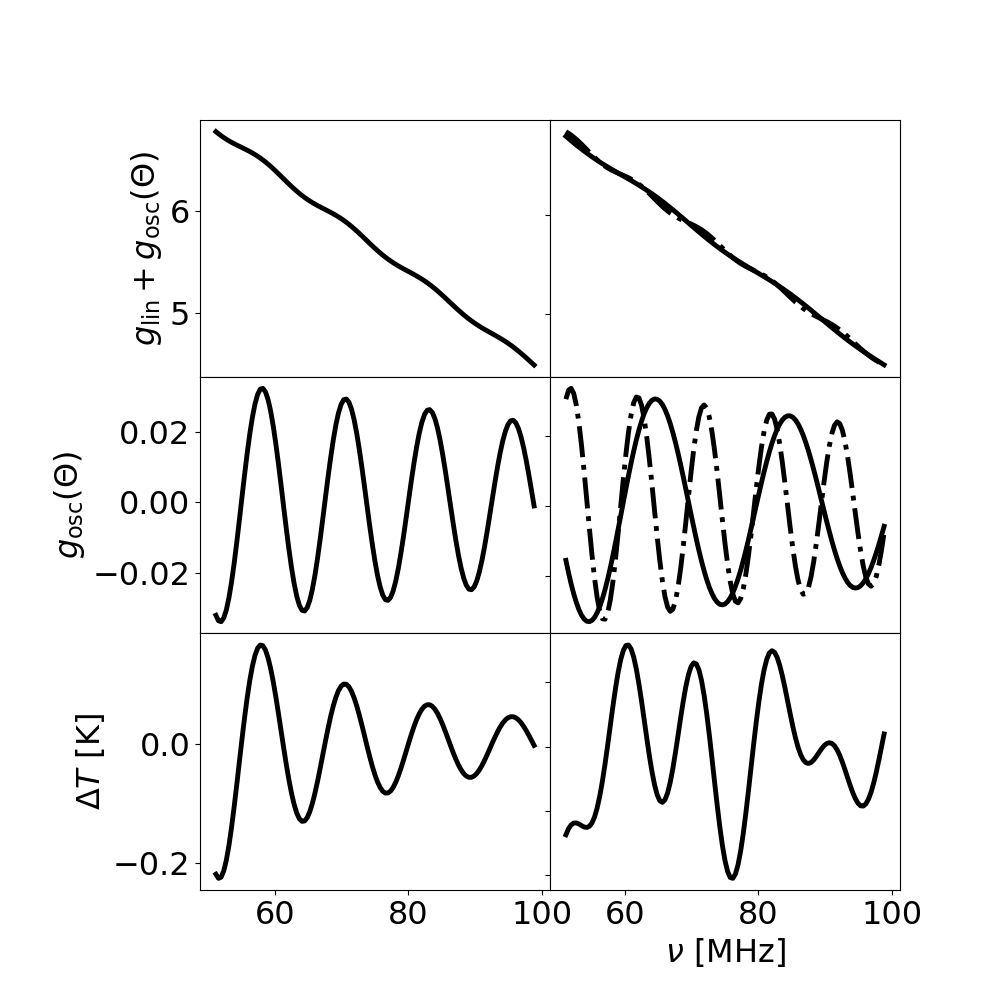}
	}
\caption{Example of calibration systematics introduced into the data by uncertainty on the beam gain model, in two limiting cases: [left column] error localised in angle from zenith and [right column] error dispersed over angle from zenith (see main text for details). [Top, left] Mock antenna gain curve, $g_\mathrm{true} = g_\mathrm{lin} + g_\mathrm{osc}$, comprised of a linear component, $g_\mathrm{lin}$, and an oscillatory component, $g_\mathrm{osc}$, designed to approximate the components of the antenna gain model for the EDGES low-band antenna, identified in B18. [Top, right] As top, left, but for ten $g_\mathrm{osc}$ with period and phase dependent on angle from zenith (for visual clarity, the two $g_\mathrm{true}$ with the maximum and minimum period $g_\mathrm{osc}$, only, are shown). [Middle, left] The oscillatory component, $g_\mathrm{osc}$, of the error localised in angle from zenith gain model. [Middle, right] The maximum and minimum period $g_\mathrm{osc}$ in the error dispersed over angle from zenith model. [Bottom] The calibration systematics, $T_\mathrm{cal}$, that would be introduced into an EDGES-like data set by a 1\% error on the amplitude of the oscillatory component(s) of the gain model, during calibration, in the two cases.}
\label{CalSys}
\end{figure}

In \autoref{CalSys}, we show a simple proof of concept example of the calibration systematics that can be introduced into the data due to low level uncertainty on the model for the beam gain. The plots in the left and right columns illustrate the cases that the beam model error is localised and dispersed in angle from zenith, respectively. In both cases, the top panel shows mock `true' antenna gain curves plotted as a function of frequency. We approximate the frequency dependence of the gain of the EDGES beam as being comprised of two components: a linear gain component and an oscillatory component designed to approximate the undulating component of the gain B18 identify following subtraction of a 5th order polynomial. We assume the `true' underlying gain for the mock antenna is given by:
\begin{equation}
\label{Eq:MockCalSysTrue}
g_\mathrm{true} = g_\mathrm{lin} + g_\mathrm{osc}.
\end{equation}
We expect any contribution to calibration systematics deriving from the structure of the non-undulating component of the gain to be subdominant. Therefore, for simplicity, we assume the linear component is independent of angle from zenith. As a proxy for the undulating component of the gain seen in B18, we consider two cases: \begin{enumerate}\item In the left column of \autoref{CalSys}, we use a single hour-angle independent undulating gain component of the form $g_\mathrm{osc} = A(\nu)\sin(2\pi*\nu/P+ \phi)$, with $A(\nu) = g_\mathrm{lin}/200$, such that the amplitude of the oscillatory component is 0.5\% of the amplitude of the linear gain component as a function of frequency; $P = 12.5~\mathrm{MHz}$ is chosen to approximate the periodicity of the undulating component of the gain at the 3 dB point in the beam model shown in B18, and $\phi=\pi/5$ is an arbitrary phase. \item In the right column of \autoref{CalSys}, we consider the case that the period and phase of the undulating gain component are a function of angle from zenith and approximate this dependence using ten $g_\mathrm{true}$ curves, with varying period and phase, such that $g_\mathrm{osc} = A(\nu)\sin(2\pi*\nu/P+ \phi)$, with $A(\nu)$ as above, and $P$ and $\phi$ are randomly drawn from Gaussian distributions with means $15~\mathrm{MHz}$ and  $\pi/5$ and standard deviations $5~\mathrm{MHz}$ and $\pi/4$, respectively. The range of phase angles is chosen to roughly reflect the relative phase between the undulating feature at zenith and the 3 dB point shown in B18, but we have confirmed that the specific choice of values shown does not fundamentally influence the result.
\end{enumerate}

The middle panel of \autoref{CalSys} shows, on the left, $g_\mathrm{osc}$ for a localised angle from zenith gain error  and, on the right, the minimum and maximum period of the ten $g_\mathrm{osc}$ used to model the gain errors dispersed in angle from zenith.

We consider the effect of a 1\% error in the amplitude of the oscillatory component of the gain (this corresponds to a 1 part in $10^{4}$ error in the amplitude of the total gain). We write our imperfect model for the gain of the mock antenna as,
\begin{equation}
\label{Eq:MockCalSysModel}
g_\mathrm{model} = g_\mathrm{lin} + 1.01g_\mathrm{osc} \ ,
\end{equation}
with $g_\mathrm{lin}$ and $g_\mathrm{osc}$ defined as in \autoref{Eq:MockCalSysTrue}. We estimate the systematic spectral structure such an error in the gain model will introduce into the imperfectly calibrated data as, 
\begin{equation}
\label{Eq:MockCalData}
T_\mathrm{cal} = T_\mathrm{sky} \sum_i\left(\dfrac{g_{\mathrm{true},i}}{g_{\mathrm{model},i}}-1\right) \ ,
\end{equation}
where $T_\mathrm{sky}$ represents the brightness temperature of the sky, which, for this mock example, we approximate with a power law fit to the EDGES data. The resulting $T_\mathrm{cal}$ calibration systematics introduced through these modelling errors are shown in the bottom row of \autoref{CalSys}. $T_\mathrm{cal}$ resulting from an error in the gain amplitude at a localised angle from zenith is well described by a power law damped sinusoid, with the frequency dependence of the damping determined by the spectral dependence of $T_\mathrm{sky}$ and $g_\mathrm{osc}$. In the case of gain errors dispersed over angle from zenith, the systematic structure is a superposition of multiple power law damped sinusoids. In both cases, for a hypothetical 1\% error in the amplitude of the undulating component of the gain model, the calibration systematics produced are of order $200~\mathrm{mK}$ in amplitude and, thus, in this mock scenario would be statistically significant features relative to the order $10~\mathrm{mK}$ noise level in the publicly available EDGES low-band spectrum.

B18 state that simulated observations with their beam model yield
residuals of $15~\mathrm{mK}$ in a five-term fit over the frequency range 52--97 MHz at $\mathrm{GHA}=10$ and residuals of $100~\mathrm{mK}$ at $\mathrm{GHA}=0$. If we fit a 5th order polynomial to the mock calibration systematics shown in \autoref{CalSys}, the RMS residuals are $\sim 80~\mathrm{mK}$, and, thus, are comparable to these values. Nevertheless, we emphasize that the use of a 1\% error in the undulating component of the gain used in \autoref{CalSys} was arbitrary and our intent in this section is only to illustrate the type of structures that can be present in the data in the case of imperfect calibration. In \autoref{ModelEvidences}, when we fit for models of systematic effects, such at those shown in \autoref{CalSys}, the amplitude of the model is a free parameter to be determined in the analysis.

We note that the presence, in the EDGES data, of effects similar to those described above is inevitable at some amplitude level. Furthermore, if there are beam model errors dispersed over angle from zenith, but larger at a particular angle, statistically significant contamination of both types described above may be present. However, it is also possible that the amplitudes of these systematic effects could be below the noise level, which will be the case if the beam is modelled sufficiently accurately. Thus, without a priori knowledge of the accuracy of the beam modelling, an important element of the analysis is to determine whether or not effects of this type are statistically significant features of the data.

\subsubsection{B18 cross-checks on the impact of the beam chromaticity correction}
\label{EDGESBeamChromaticityCorrectionB18CrossChecks}

B18 vary the beam model in their chromaticity correction and look at the impact this variation has on the amplitude of the recovered signal. Specifically, they fit the depth of the absorption trough for four types of beam chromaticity correction: \begin{enumerate*} \item no beam correction, \item no beam correction with the spectral range of the fit restricted to 65--95 MHz, \item using a beam correction model derived from an HFSS EM simulation and \item using a beam correction model derived from a FEKO EM simulation.\end{enumerate*}
%
In these cases, the depth of the absorption trough varies between a minimum of $370~\mathrm{mK}$ (when using no beam correction and fitting over the full $50-100~\mathrm{MHz}$ spectral band) and $670~\mathrm{mK}$ (when using the HFSS EM simulation of the beam).

However, this comparison ignores the possibility that other components in the model could be simultaneously modified and actually produce a better fit to the data. Additionally, a statistically robust metric for model comparison is not given with the results. This limits what can be inferred from the comparison.

For example, it is very likely that when no beam correction is applied, fitting a fifth order polynomial model to describe the foregrounds, ionosphere and calibration systematics (which in this case will certainly be in the data), plus a global signal model, will fail to accurately describe the data. Instead, a higher order polynomial that can better model the calibration systematics, or an explicit model for the calibration systematics in the data, or both, will be preferred. If such a model, that better described the data, was fit, and the amplitude of an unrestricted flattened Gaussian model was jointly estimated, there is no a priori reason to expect the amplitude of this flattened Gaussian to be similar to that found by B18. The same argument equally holds true if the chromaticity correction is insufficient to reduce calibration systematics to below the noise in the primary analysis in B18, which uses a model with no component to describe systematic effects.

Ideally, a more comprehensive range of plausible models should be fit to the data, accounting for the possibility of calibration systematics and allowing the full range of free parameters of each model to vary when fitting. The amplitude of the global signal in the models that best describe the data provides then a robust measure of the level of constraint on the amplitude of the signal that can be obtained from the data\footnote{We note that the inclusion of model components to fit plausible calibration systematics is not an automatic rejection of the B18 assertion that these errors are small and do not influence their result. Rather, if this is the case, we expect the Bayesian evidence to favour models that do not include these components, and we test for this eventuality in our analysis.}. We carry out such an analysis in this paper and, as such, we explicitly incorporate model components for systematic effects in our data model and use Bayesian model selection to determine whether these are preferred or disfavoured relative to those excluding these components. Next, we briefly outline the principles of Bayesian inference on which this analysis is based.

\subsection{Bayesian inference}
\label{BayesianInference}

Bayesian inference provides a consistent approach to estimate a set of parameters, $\sTheta$, from a model, $M$, given a set of data $\bm{D}$ and, through the use of the Bayesian evidence $\mathrm{Pr}(\bm{D}\vert M_{i})\equiv\mathcal{Z}_{i}$, to select from a set of models the ones that best describe the data. Bayes' theorem states that,
\begin{equation}
\label{Eq:BayesEqn}
\mathrm{Pr}(\sTheta\vert\bm{D},M) = \dfrac{\mathrm{Pr}(\bm{D}\vert\sTheta,M)\ \mathrm{Pr}(\sTheta\vert M)}{\mathrm{\mathrm{Pr}}(\bm{D}\vert M)} = \dfrac{\mathcal{L}(\sTheta)\pi(\sTheta)}{\mathcal{Z}}, 
\end{equation}
where $\mathrm{Pr}(\sTheta\vert\bm{D},M)$ is the posterior probability distribution of the parameters, $\mathrm{Pr}(\bm{D}\vert\sTheta,M) \equiv \mathcal{L}(\sTheta)$ is the likelihood and $\mathrm{Pr}(\sTheta\vert M) \equiv \pi(\sTheta)$ is the prior probability distribution of the parameters.

The Bayesian evidence (the factor required to normalise the posterior over the parameters), is given by,
\begin{equation}
\label{Eq:Evidence}
\mathcal{Z}=\int\mathcal{L}(\sTheta)\pi(\sTheta)\mr{d}^{n}\sTheta,
\end{equation}
where $n$ is the dimensionality of the parameter space. Comparison of the evidence for different models enables a statistically robust selection of a preferred model for the data. As the average of the likelihood over the prior, the evidence is larger for a model if more of its parameter space is likely and smaller for a model where large areas of its parameter space have low likelihood values, even if the likelihood function is very highly peaked.  Thus, the evidence automatically implements Occam's razor: a simpler theory with a compact parameter space will have a larger evidence than a more complicated one, unless the latter is significantly better at explaining the data.

Bayesian inference addresses model comparison between two possible models for a data set, $M_{0}$ and $M_{1}$, via consideration of $R$, the ratio of posterior odds in favour of the hypothesis and the prior odds in favour of the hypothesis,
\begin{equation}
\label{Eq:BayesFactor}
R_{10} = \dfrac{\mathrm{Pr}(M_{1}\vert\bm{D})}{\mathrm{Pr}(M_{0}\vert\bm{D})} = \dfrac{\mathrm{Pr}(\bm{D}\vert M_{1})\mathrm{Pr}(M_{1})}{\mathrm{Pr}(\bm{D}\vert M_{0})\mathrm{Pr}(M_{0})} = B_{10}\dfrac{\mathrm{Pr}(M_{1})}{\mathrm{Pr}(M_{0})} \ ,
\end{equation}
where the ratio of posterior odds in favour of the hypothesis is called the `Bayes Factor', $B$, $\mathrm{Pr}(\bm{D}\vert M_{1}) \equiv \mathcal{Z}_{1}$ and $\mathrm{Pr}(\bm{D}\vert M_{0}) \equiv \mathcal{Z}_{0}$ are the marginal likelihoods of the data in models $M_{1}$ and $M_{0}$, respectively, and $\mathrm{Pr}(M_{1})/\mathrm{Pr}(M_{0})$ is the prior probability ratio for the two models set before any conclusions have been drawn from the data set. In this work, we assume a priori that the different models we consider are equally likely, in which case, $R=B$.

The Bayes factor is a summary of the evidence provided by the data for one model, as opposed to another. Different classification schemes exist for interpreting the significance that is implied by a given Bayes factor. In this paper, when comparing models for the data, we follow \citet{KandR}, who consider $\ln({B_{0i}}) = \ln(\mathcal{Z}_{0}) -\ln(\mathcal{Z}_{i}) = 3$, corresponding to a ratio of posterior odds of $\sim20$, to constitute strong evidence in favour of $M_{0}$ over $M_{i}$.

\subsection{Signal and noise models}
\label{EDGESDataModel}

We can write the EDGES low-band data set as a vector $\mathbfit{d} = \mathbfit{s} + \sdelta\mathbfit{n}$, where the signal $\mathbfit{s}$ is a maximum likelihood estimate of the intrinsic brightness temperature spectrum of the sky, in kelvin, in the region observed by EGDES, and $\sdelta\mathbfit{n}$ is the noise on the data.

We define a Gaussian likelihood function for our spectral model of the EDGES low-band as,
\begin{eqnarray}
\label{Eq:BasicVisLike}
\mathcal{L}(\sTheta) \propto \frac{1}{\sqrt{{\mathrm{det}(\mathbfss{N})}}} \exp\left[-\frac{1}{2}\left(\mathbfit{d} - \mathbfit{m}(\sTheta)\right)^{\dagger}\mathbfss{N}^{-1}\left(\mathbfit{d} - \mathbfit{m}(\sTheta)\right)\right],
\end{eqnarray}
where $\mathbfit{m}(\sTheta)$ is our model for the signal, described  in detail below, and we model the noise on the spectrum as an uncorrelated Gaussian random field, with covariance matrix $\mathbfss{N}$. The elements of the covariance matrix are given by $\mathbfss{N}_{ij} = \left< n_in_j^*\right> = \delta_{ij}f_{N}(\sigma_{j})$, where $f_{N}(\sigma_{j})$ is a model of the variance of the noise on the data that we will discuss further in \autoref{CovarianceMatrixModel}, $\left< ... \right>$ represents the expectation value and $\sigma_{j}$ denotes the noise expectation in element $j$ of the spectrum.

Following H18, we start by defining a physical model of the intrinsic sky signal as,
\begin{equation}
\label{Eq:SkyModelPhysical}
\mathbfit{m} = (\mathbfit{T}_\mathrm{\bf{R}} + \sdelta \mathbfit{T}_\mathrm{\bf{b}} + \mathbfit{T}_\mathrm{\bf{Fg}}) \circ e^{-\bm\tau_\mathrm{\bf{ion}}} + T_\mathrm{e}(1 - e^{-\bm\tau_\mathrm{\bf{ion}}}) \ . 
\end{equation}
Here, $\mathbfit{T}_\mathrm{\bf{R}}$ is as defined in \autoref{Eq:DifferentialBrightness}, $\circ$ denotes the Hadamard `element-wise' product of vectors, $\sdelta \mathbfit{T}_\mathrm{\bf{b}}$ is the vectorised differential brightness temperature between the hydrogen spin temperature and the background temperature, as defined in \autoref{Eq:DifferentialBrightness}, $\mathbfit{T}_\mathrm{\bf{Fg}}$ is the brightness temperature of foreground emission from the Galaxy and extragalactic sources. $\bm\tau_\mathrm{\bf{ion}} = \tau_0 (\bm\nu/\nu_{c})^{-2}$ is the opacity of the ionosphere for the vector of channel frequencies, $\bm\nu$, in the EDGES data set, with $\tau_0$ the opacity at reference frequency $\nu_{c}$ and the quadratic frequency dependence of $\bm\tau_\mathrm{\bf{ion}}$ is taken from the measurements of \citet{2015RaSc...50..130R}. Finally, $T_\mathrm{e}$ is the temperature of the electrons in the ionosphere.

The opacity of the ionosphere in the frequency range of the data is of order $\tau_\mathrm{ion} \sim 10^{-2} << 1$, thus \autoref{Eq:SkyModelPhysical} is well approximated by\footnote{We have confirmed that the term $\sdelta \mathbfit{T}_\mathrm{\bf{b}}\bm\tau_\mathrm{\bf{ion}}$ has a negligible effect on recovery of the signal and we have therefore excluded it from the model.},
\begin{equation}
\label{Eq:SkyModelApprox}
\mathbfit{m} = \sdelta \mathbfit{T}_\mathrm{\bf{b}} + (\mathbfit{T}_\mathrm{\bf{R}} + \mathbfit{T}_\mathrm{\bf{Fg}}) \circ (1-\bm\tau_\mathrm{\bf{ion}}) + T_\mathrm{e}\bm\tau_\mathrm{\bf{ion}} \ .
\end{equation} 

Additionally, to recover unbiased estimates of the aforementioned parameters, it is essential that any systematic effects resulting from imperfect calibration of the data are modelled. We, therefore, incorporate models for calibration systematics and write our final model vector as, 
\begin{equation}
\label{Eq:SkyModelApprox2}
\mathbfit{m} = \sdelta \mathbfit{T}_\mathrm{\bf{b}} + \bar{\mathbfit{T}}_\mathrm{\bf{Fg}} + \mathbfit{T}_\mathrm{\bf{cal}} \ ,
\end{equation}
where, $\bar{\mathbfit{T}}_\mathrm{\bf{Fg}} = (\mathbfit{T}_\mathrm{\bf{R}} + \mathbfit{T}_\mathrm{\bf{Fg}}) \circ (1-\bm\tau_\mathrm{\bf{ion}}) + T_\mathrm{e}\bm\tau_\mathrm{\bf{ion}} + \bar{\mathbfit{T}}_\mathrm{\bf{cal}}$ is a model for a combination of foregrounds and systematic effects that obscure the global 21 cm signal of interest, and $\mathbfit{T}_\mathrm{\bf{cal}}$ and $\bar{\mathbfit{T}}_\mathrm{\bf{cal}}$ encode calibration systematics localised or dispersed in angle from zenith, respectively, as detailed in \autoref{EDGESData}. In the remainder of this section we describe our model parametrisations for $\sdelta \mathbfit{T}_\mathrm{\bf{b}}$, $\bar{\mathbfit{T}}_\mathrm{\bf{Fg}}$, $\mathbfit{T}_\mathrm{\bf{cal}}$ and the noise on the data.

\subsubsection{$\sdelta T_\mathrm{b}$ parametrisation}
\label{Global21cmModel}

We consider three model parametrisations for the global 21 cm signal profile: \begin{enumerate}
\item \textit{Flattened Gaussian -} to facilitate direct comparison with B18, we take as our first global signal model, a flattened Gaussian profile of the form,
\begin{equation}
T_\mathrm{b}(\nu) = -A\left(\frac{1-e^{-\tau e^B}}{1-e^{-\tau}}\right) \ ,
\label{Eq:FlattenedGaussian}
\end{equation}
where,
\begin{equation}
B = \frac{4(\nu-\nu_0)^2}{w^2}\log\left[-\frac{1}{\tau}\log\left(\frac{1+e^{-\tau}}{2}\right)\right] \ ,
\label{Eq:FlattenedGaussianB}
\end{equation}
and $A$, $\nu_0$, $w$ and $\tau$ describe the amplitude and central frequency, width and flattening of the absorption trough, respectively. 
\item \textit{Gaussian -} H18 find that, when the presence of a sine wave systematic in the data model is subtracted and a polynomial foreground model plus flattened Gaussian profile are fit to the residuals, the flattening factor in the best fit of the flattended Gaussian profile tends towards low values and a comparably good fit to the data is obtained by substituting \autoref{Eq:FlattenedGaussian} for a simple three parameter Gaussian model for the signal of the form,
\begin{equation}
T_\mathrm{b}(\nu) = -Ae^{\left[\frac{-4\log(2)(\nu-\nu_{0})^{2}}{w^{2}}\right]} \ .
\label{Eq:StandardGaussian}
\end{equation}
Additionally, S19 find that, in a joint fit for a sine wave systematic, a maximally smooth polynomial foreground model (see \citealt{2015ApJ...810....3S} for details) and a three parameter Gaussian global signal model, the best fitting global signal model has an amplitude $A = 133 \pm 60~\mathrm{mK}$ that lies below the $209~\mathrm{mK}$ maximum depth possible in models following standard cosmological assumptions. Motivated by these findings and by the fact that the evidence will favour a simpler three parameter Gaussian parametrization if it can provide a comparably good fit to the four parameters model given in \autoref{Eq:FlattenedGaussian}, we use \autoref{Eq:StandardGaussian} as our second global signal model parametrization.
\item \textit{{\sc{ARES}} simulation -} Neither of the model parametrisations given by \autoref{Eq:FlattenedGaussian} or \autoref{Eq:StandardGaussian} describe the physics that creates the 21 cm absorption profile. Rather, they are simply functional forms that it is hoped can provide reasonable descriptions of the intrinsic shape of a cosmological absorption profile in the data. An advantage of these simple parametrisations is that their absorption depths are constrained only by the priors we choose to place on them. As such, allowing for the presence of a deeper than expected feature in the data is trivial. However, two notable deficiencies of these model parametrisations are that they have enforced symmetry about $\nu_{0}$ by construction and that they are limited to modelling either absorption or emission. In contrast, the rate at which Ly$\alpha$ photons couple $T_\mathrm{K}$ to $T_\mathrm{\gamma}$ via the Wouthuysen--Field effect is not guaranteed to match the subsequent rate of heating of the IGM by $T_\mathrm{K}$. Thus, an asymmetric absorption is, in general, expected. Further, if $T_\mathrm{K}$ is heated above $T_\mathrm{\gamma}$ in the frequency range of the EDGES low-band, then it will be accompanied by a subsequent emission feature, which will not be modelled by Gaussian or flattened Gaussian profiles, potentially biasing recovered estimates.

To address these limitations, we construct and use an atlas of {\sc{ARES}} simulations as our model global 21 cm signals. Here, we give a briefly outline of the approach taken by {\sc{ARES}} for calculating the global 21 cm signal. For a detailed description of the {\sc{ARES}} simulations, see \citet{2014MNRAS.443.1211M}.

{\sc{ARES}} calculates $\sdelta T_\mathrm{b}$ using \autoref{Eq:DifferentialBrightness}, under the approximation that fluctuations deriving from spatial variation of the baryon overdensity field and from redshift space distortions sourced from the gradient of the proper velocity along the line of site can be neglected. In this approximation, to calculate $\sdelta T_\mathrm{b}$ one must first derive the hydrogen neutral fraction (or, equivalently, the volume-filling factor, $Q_{H_\mathrm{II}} = 1 - x_{\ion{H}{I}}$), the background radiation temperature and the hydrogen spin temperature. The hydrogen spin temperature, itself, is a function of the coupling coefficients for collisions (a function of the fraction of free electrons in the IGM, $x_\mathrm{e}$), the UV scattering (a function of the Ly$\alpha$ intensity, $J\alpha$) and the kinetic temperature of the hydrogen gas.

{\sc{ARES}} employs a two phase model for the IGM: a fully ionized phase ($H_\mathrm{II}$ regions; which is dark at redshifted 21 cm wavelengths and whose sole characteristic is its volume-filling factor, $Q_{H_\mathrm{II}}$) and a mostly neutral `bulk IGM' phase, from which the 21 cm signal derives.

The time evolution of the IGM's properties, $Q_{H_\mathrm{II}}$, $T_\mathrm{K}$, and $x_\mathrm{e}$, is governed by the properties of sources in the volume. A model for the volume-averaged emissivity of sources as a function of time is constructed under the assumption that it is proportional to the product of the rate of matter collapse into sources in haloes above a minimum virial temperature threshold, $T_\mathrm{min}$, the bolometric luminosity density of the sources and their spectral energy density (SED). The first of these properties can be expressed as functions of the star formation efficiency (i.e. the fraction of collapsing gas that becomes stars), $f_\mathrm{\star}$, and the last two can be expressed as functions of the normalisation of the X-ray luminosity to star formation rate (LX-SFR) relation at high redshift, $f_\mathrm{X}$, and the number of ionizing photons, $N_\mathrm{ion}$, and Lyman-Werner photons, $N_\mathrm{lw}$, emitted per baryon of star formation. With these specified, the corresponding radiation field is evolved to obtain the mean meta-galactic background intensity, $J_\nu$, as a function of redshift. Using $J_\nu$, the ionization and heating rates are derived, and the state of the gas in each phase of the two-zone IGM ($Q_{H_\mathrm{II}}$, $x_\mathrm{e}$, $T_\mathrm{K}$) is calculated. Finally, given $Q_{H_\mathrm{II}}$, $x_\mathrm{e}$, $T_\mathrm{K}$ and $J_\nu$, $\sdelta T_\mathrm{b}$ is derived, as a function of time, via \autoref{Eq:DifferentialBrightness}.

The global 21 cm signal is strongly influenced by the five parameters ($T_\mathrm{min}$, $f_\mathrm{\star}$, $f_\mathrm{X}$, $N_\mathrm{ion}$ and $N_\mathrm{lw}$) described above. {\sc{ARES}} allows for $f_\mathrm{\star}$ to be parametrised as a function of halo mass and for the UV emissivity of galaxies to be determined via a choice of stellar population synthesis spectra; however, in this work, our primary goal, with respect to using {\sc{ARES}} global 21 cm signals, is to overcome the limitations described earlier by constructing an atlas of models that spans the global 21 cm signal space in the standard cosmology limit (i.e. with $T_\mathrm{R} = T_{\gamma}$ at CD, and cooling of the hydrogen gas following decoupling from $T_{\gamma}$ that occurs exclusively as a result of adiabatic expansion). We therefore focus on the simplest {\sc{ARES}} simulation scheme for constructing such an atlas, with: $f_\mathrm{\star}$ a constant and $N_\mathrm{ion}$ and $N_\mathrm{lw}$ independently sampled from.

We construct an atlas of 100,000 models derived from {\sc{ARES}} simulations that span the space of global 21 cm signal shapes in the standard cosmology limit. We parametrise the models by the astrophysical parameters described above, using prior bounds\footnote{The prior ranges are selected to encompass a liberal range of parameter values sufficient to derive simulations spanning the space of global 21 cm signal shapes in the standard cosmology limit.} on the parameters, as listed in \autoref{Tab:AstroModelPriors}. Within the prior ranges, we sample 10 values of each parameter, with $T_\mathrm{min}$, $N_\mathrm{lw}$ and $N_\mathrm{ion}$ values sampled uniformly in log10 and $f_\mathrm{X}$ and $f_\mathrm{\star}$ values uniformly sampled in linear space, between their respective prior minima and maxima. By default, {\sc{ARES}} simulations output $\sdelta T_\mathrm{b}$ as a function of redshift (or, equivalently, frequency) over a larger spectral range and higher spectral resolution than the publicly available EDGES low-band data. We, therefore, interpolate the simulation outputs to the central frequencies of the EDGES low-band data set to produce our final atlas of models.
\end{enumerate}

\begin{table}
\caption{Minimum and maximum of the priors on the {\sc{ARES}} parameters considered in the analysis. Parameter definitions are given in the main text.}.
\centerline{
\begin{tabular}{l l l l }
\toprule
Coefficient & Parameter & Prior Min & Prior Max   \\
\midrule
$C_{0}$ & $f_\mathrm{X}$ & $1.0$ & $100.0$ \\
$C_{1}$ & $f_{\star}$ & $0.1$ & $1.0$ \\
$C_{2}$ & $T_\mathrm{min}$ & $10^{3}$ & $10^{5}$ \\
$C_{3}$ & $N_\mathrm{lw}$ & $9690.0$ & $10^{5}$ \\
$C_{4}$ & $N_\mathrm{ion}$ & $4000.0$ & $10^{5}$ \\
\bottomrule
\end{tabular}
}
\label{Tab:AstroModelPriors}
\end{table}

\subsubsection{$\bar{T}_\mathrm{Fg}$ parametrisation}
\label{FgModel}

$\bar{T}_\mathrm{Fg}$ is a general model to fit structure in the data imprinted by astronomical foregrounds, the ionosphere and calibration systematics of the form expected from beam model errors dispersed over angle from zenith (if present; see \autoref{EDGESData}).

It has been shown by e.g. \citet{2015ApJ...799...90B} and \citet{2016MNRAS.461.2847B} that realistic models of foreground emission can be well described by a log-polynomial foreground model. Furthermore, \citet{2016MNRAS.461.2847B} show that accurate estimates of the parameters of the global 21 cm signal are recovered when jointly estimated with a log-polynomial foreground model of intermediate (3rd to 5th) order, with the specific order required dependent on instrumental details.

Ionospheric absorption introduces a spectral dependence via $\tau_\mathrm{ion}$ that is reasonably described as quadratic. As such, a log-polynomial model will also provide a good fit to the combination of intrinsic foregrounds and ionospheric effects.

Finally, while great effort has been made to calibrate the EDGES low-band data set, \citet{2018Natur.564E..35B} state that the foreground model used in the analysis of B18 is intended to account for a combination of astronomical foregrounds, ionospheric effects and any residual calibration effects, and, as described in \autoref{EDGESData}, small mismatches in a combination of the amplitude, phase and period of the undulating gain feature present in the EDGES electromagnetic beam model and the true instrumental transfer function will produce a degenerate combination of sky temperature scaled sinusoidal systematics \footnote{Additionally, \citet{2019arXiv190805303S} have recently shown that polarised foregrounds may also have imparted spectrally fluctuating structure to the EDGES low-band data.} that will impart a more general fluctuating systematic structure to the data (e.g. see \autoref{CalSys}, bottom right). A log polynomial model with sufficient degrees of freedom to model structure on the scales present, provides a compact parametrisation for such structure.

We, therefore, model the combination of these components using a log-polynomial parametrisation of the form,
\begin{equation}
\label{Eq:LogPoly}
\bar{T}_\mathrm{Fg} = 10^{\sum_{i=1}^{n} d_{i}\log_{10}(\nu/\nu_{0})^i } \ ,
\end{equation}
where $d_{i}$ are the coefficients to be fit for, and we use a reference frequency $\nu_{0} = 75~\mathrm{MHz}$, corresponding to the central frequency of the EDGES low-band.

The combination of power law foreground emission and ionospheric effects are expected to be well described by intermediate-order log-polynomials (e.g. H18). If there is only intrinsic foreground and ionospheric structure in the data, models with low to intermediate order log-polynomials will be preferred and the evidence will disfavour models with unnecessary higher order terms. When performing our model selection analysis, we, therefore, test whether models with log-polynomial components beyond 5th order better describe the data and use this as a proxy for the presence of additional oscillatory structure in the data not attributable to the foregrounds, ionosphere, or global 21 cm signal.

We note that maximally smooth polynomials provide a potential alternate choice of foreground model (see S19). However, fluctuating spectral features such as those that could be introduced by calibration systematics and polarised foreground emission cannot be modelled with the maximally smooth subset of log polynomials and, thus, will be absorbed by the remaining model components, potentially biasing the recovered signal estimates. We, therefore, do not impose this constraint on our model, since, if such structures are present, the reduction in bias facilitated by the use of a combined foreground plus systematic model parametrised by a general log polynomial model, relative to a maximally smooth component, is preferable to reducing the covariance between the model components\footnote{We also validated this choice computationally by replacing the log-polynomial component of the highest Bayesian evidence data models found in \autoref{ModelEvidences} with equal order maximally smooth polynomials. In all cases, we find that the models incorporating unconstrained log-polynomials are decisively preferred ($\log(B)>3$).}.

\subsubsection{$T_\mathrm{cal}$ parametrisation}
\label{CalibrationSystematicModel}

Ideally, multiplicative model(s) for the antenna gain and its uncertainty would be included in \autoref{Eq:SkyModelPhysical} and \autoref{Eq:SkyModelApprox2} and  marginalised over when estimating the global 21 cm signal, or the best fitting antenna gain model would be determined as a component of a Bayesian model selection analysis. However, such an approach requires using the raw EDGES data, before it is averaged to a single spectrum, as is the case in the publically available data. Fitting for an further additive component of the data model, designed to capture the effect of mismodelling the gain, provides an alternate approach that can be applied to the publicly released spectral data.

H18 and S19 find that models incorporating a sinusodal feature with a $12~\mathrm{MHz}$ period and a $\sim0.06~\mathrm{mK}$ frequency independent amplitude can provide a good fit to the data. Motivated by the calibration discussion in \autoref{EDGESData}, we generalise this parametrisation and use a power law damped sinusoid to model a systematic feature that would be expected to result from a systematic error deriving from a gain calibration model residual dominated by an error in the amplitude of the undulating component of the gain at a single angle from zenith (this is in addition to the log polynomial model for calibration errors deriving from errors in the amplitude of the undulating component of the gain across a broader range of angles from zenith, described in \autoref{FgModel}). We write this model as\footnote{Our prior range of the power law damping envelope of the sinusoid encloses zero, therefore, our generalised systematic model also encompasses the flat sinusoidal systematic models considered by H18 and S19.},
\begin{equation}
\label{Eq:PLDampedSinusoid}
 T_\mathrm{cal} = (\nu/\nu_{0})^{b} [a_{0}\sin(2\pi\nu/P) + a_{1}\cos(2\pi\nu/P)] \ ,
\end{equation}
where $b$ is the power law damping index, $P$ is the period of the sinusoid in $\mathrm{MHz}$ and $a_{0}$ and $a_{1}$ are the amplitudes of the sine and cosine components of the systematic model in $\mathrm{K}$, respectively.

In our analysis, we calculate the Bayesian evidence of models incorporating log-polynomials with orders between 3 and 10, in combination with the explicit calibration model described above. Covariance between these model components has the potential to impart a strong dependence between the log-polynomial order and the amplitudes of the systematic components in \autoref{Eq:PLDampedSinusoid}, required to optimally describe the data. Therefore, to ensure the analysis is not prior bound limited, irrespective of the log-polynomial order, we parametrise the amplitudes as $a_{0} = 10^{\bar{a}_{0}}$ and $a_{1} = 10^{\bar{a}_{1}}$, respectively, and sample uniformly from $\bar{a}_{0}$ and $\bar{a}_{1}$ when incorporating a damped sinusoid systematic model component.

\subsubsection{Covariance matrix, $\mathbfss{N}$, parametrisation}
\label{CovarianceMatrixModel}

The ideal radiometer equation for the noise in observations of length $\tau$ and channel width $\Delta\nu$, using an antenna with a system temperature $T_\mathrm{sys} = T_\mathrm{sky} + T_\mathrm{receiver}$, with $T_\mathrm{sky}$ and $T_\mathrm{receiver}$ the sky and receiver temperatures, respectively, is given by,
\begin{equation}
\label{Eq:IdealRadiometer1}
 \sigma_\mathrm{T} = \dfrac{T_\mathrm{sys}}{\sqrt{\Delta\nu \tau}} \ .
\end{equation}
For the EDGES dipole, $T_\mathrm{receiver} \approx 200~\mathrm{K}$ (see B18 Extended Data Figure 5). For a calibrated data set: $T_\mathrm{sky} \simeq d$. The publicly available EDGES low-band data has a sky observation time of $\tau_\mathrm{sky} = 428~\mathrm{h}$, an effective integration time\footnote{The EDGES data acquisition system has a duty cycle of about 50\% and a spectral window function efficiency of about 50\%, thus, yielding an effective integration time that is approximately 25\% of the time observing the sky (B18 Extended Table 1).}, $\tau_\mathrm{eff} = \tau_\mathrm{sky}/4 = 107~\mathrm{h}$ (see B18 Extended Table 1) and a channel width $\Delta\nu = 390.625~\mathrm{kHz}$. As such, the intrinsic noise level in the publicly available EDGES data is well approximated by,
\begin{equation}
\label{Eq:IdealRadiometer2}
 \sigma_\mathrm{T} = \dfrac{T_\mathrm{sys}}{3.879\times10^5} \ .
\end{equation}

The sky temperature at the low frequency end of the data set is $T_\mathrm{sky}(51.2~\mathrm{MHz}) \simeq 4.6\times10^{3}~\mathrm{K}$, and at the high end it is $T_\mathrm{sky}(98.8~\mathrm{MHz}) \simeq 8.6\times10^{2}~\mathrm{K}$. Thus, from \autoref{Eq:IdealRadiometer1}, we expect noise in the data to be comprised of a low level ($\sim0.5~\mathrm{mK}$) approximately white component due to the receiver and a sky noise component that, like $d$, scales as a power law with frequency, between a maximum and minimum of $\sigma_\mathrm{T}(51.2~\mathrm{MHz}) \simeq 12.4~\mathrm{mK}$ and $\sigma_\mathrm{T}(98.8~\mathrm{MHz}) \simeq 2.2~\mathrm{mK}$, respectively.

We note that even the maximum anticipated noise level is approximately half the approximately $25~\mathrm{mK}$ RMS of the residuals found in the analysis of the EDGES data in B18 and the predicted noise level at the high frequency end of the band is approximately an order of magnitude lower than found in that analysis.
 
However, there are potentially at least two additional sources of noise-like structure in the data. One can be imparted by the EDGES data calibration, if there is a mismatch in the amplitude of undulating chromatic gain features seen in the electromagnetic beam model and the amplitude of corresponding features in the true instrumental transfer function that vary rapidly in frequency, or from noise-like errors in the sky model used in the chromaticity correction component of the calibration. Each can impart a sky-temperature modulated noise similar in shape to the intrinsic noise. A second is motivated by the possible presence of polarised foreground emission deriving from a range of Faraday depths (see \citealt{2019arXiv190805303S}). Structure deriving from this emission on spectral scales smaller than the channel width of the publicly available data set cannot be recovered with perfect fidelity as a component of the data model and, thus, will contribute to the effective noise level in the data.

Finally, the publicly available EDGES low-band data set has non-uniform channel weights (the fraction of data integrated in each spectral bin), with frequencies above $87~\mathrm{MHz}$, in particular, having significantly lower weights due to radio frequency interference (RFI) from FM radio stations (see B18, Extended Data Figure 7). To account for this, we use the normalised channel weights given in B18 Extended Data Figure 7, and write them as a vector, $w$.  In the remainder of this section, we use $w$ to account for the effect of the relative weights of the data in our covariance matrix parametrisations.

We, therefore, use a sufficiently general uncorrelated data covariance model to account for intrinsic radiometer noise, noise-like structure deriving from imperfect calibration and polarised foregrounds, which we model as a combination of radiometer-like and white noise-like structure. The diagonal elements of this covariance matrix are given by,
\begin{equation}
\label{Eq:CovModelScaledRadiometerPlusWN}
N_\mathrm{ii} = \dfrac{1}{w_{i}} \left[\dfrac{\alpha_\mathrm{rn} T_{\mathrm{sys},i}^{2}}{{3.879\times10^5}} + \sigma_\mathrm{wn} \right] \ .
\end{equation}
Here, the parameters $\alpha_\mathrm{rn}$ and $\sigma_\mathrm{wn}$ are fit for in the analysis; $\alpha_\mathrm{rn}$ is a scale factor of the radiometer noise in the data, with $\alpha_\mathrm{rn}=1$ corresponding to zero contribution to the effective radiometer noise from miscalibration, and $\sigma_\mathrm{wn}$ is the absolute level, in kelvin, of spectrally flat noise-like structure in the data.
We note that, from \autoref{Eq:IdealRadiometer2}, we expect $\alpha_\mathrm{rn} \ge 1$; however, when analysing the data, we set our prior range to include $\alpha_\mathrm{rn}=0$, in which case, barring the data occupancy weights\footnote{We have confirmed that while setting the data occupancy weights to unity, thus allowing for an exact match with the form of the noise model used in B18, H18 and S19, alters the evidence of individual models, it does not change the conclusions of our analysis.}, the data covariance is reduced to an intrinsic white noise model, matching that considered in B18, H18 and S19.

\section{Model Evidences}
\label{ModelEvidences}

\begin{table}
\caption{Minimum and maximum of the uniform priors on the parameters of the models used in the analysis and the model components the parameters are associated with. Parameter definitions are given in the main text.}.
\centerline{
\begin{tabular}{l l l l l }
\toprule
Model     & Parameter & Unit & Prior     & Prior       \\
component &           &      &       Min &       Max   \\
\midrule
$\bar{T}_\mathrm{Fg}$                 &  $d_{0}$  & $\log_{10}(T[\mathrm{K}])$ &  $-5.0$  &  $5.0$ \\
                                      &  $d_{1}$  & - &  $-3.0$  &  $-2.0$ \\
                                      &  $d_{2-10}$  & - &  $-100.0$  &  $100.0$ \\
\midrule
$\sdelta T_\mathrm{b}$: FG            &  $\tau$  & - &  $0$  &  $40$ \\
$\sdelta T_\mathrm{b}$: FG, G         &  $A$  & $\mathrm{K}$ &  $0.0$  &  $4.0$ \\
                                      &  $\nu_0$  & $\mathrm{MHz}$ &  $60$  &  $90$ \\
                                      &  $w$  & $\mathrm{MHz}$ &  $5$  &  $40$ \\
\midrule
$\sdelta T_\mathrm{b}$: {\sc{ARES}}          &  $f_\mathrm{X}$  & - &  $1.0$  &  $100.0$ \\
                                      &  $f_{\star}$  & - &  $0.1$  &  $1.0$ \\
                                      &  $T_\mathrm{min}$  & $\mathrm{K}$ &  $10^{3}$  &  $10^{5}$ \\
                                      &  $N_\mathrm{lw}$  & - &  $9690.0$  &  $10^{5}$ \\
                                      &  $N_\mathrm{ion}$  & - &  $4000.0$  &  $10^{5}$ \\
\midrule
$T_\mathrm{cal}$                      &  $\bar{a}_{0}$  & $\log_{10}(T[\mathrm{K}])$ &  $-10$  &  $2$ \\
                                      &  $\bar{a}_{1}$  & $\log_{10}(T[\mathrm{K}])$ &  $-10$  &  $2$ \\
                                      &  $P$  & $\mathrm{MHz}$ &  $10.0$  &  $15.0$ \\
                                      &  $b$  & - &  $-4.0$  &  $4.0$ \\
\midrule
$N$                                   &  $\alpha_\mathrm{rn}$  & - &  $0.0$  &  $20.0$ \\
                                      &  $\alpha_\mathrm{wn}$  & $\mathrm{K}$ &  $10^{-4}$  &  $10^{-1}$ \\
\bottomrule
\end{tabular}
}
\label{Tab:ModelPriors}
\end{table}

\begin{table*}
\caption{The constituent components comprising the models for the EDGES low-band data and their derived log Bayesian evidences.}.
\centerline{
\begin{tabular}{l l l l l l l }
\toprule
Model number & Global signal & Log-polynomial & Damped sinusoid, & Noise model, & log(evidence) & Residual RMS             \\
           & model, $\sdelta T_\mathrm{b}$         & order, $\bar{T}_\mathrm{Fg}$          &  $T_\mathrm{cal}$               & N          &               & [mK]                   \\
\midrule
1 & - & 3 & - & white & $-4289.96 \pm 0.36$ & 255.7 \\
2 & - & 4 & - & white & $-4121.56 \pm 0.35$ & 237.9 \\
3 & {\sc{ARES}} & 3 & - & white & $-3429.35 \pm 0.25$ & 224.4 \\
4 & {\sc{ARES}} & 4 & - & white & $-3402.66 \pm 0.25$ & 217.2 \\
5 & - & 3 & \autoref{Eq:PLDampedSinusoid} & white & $-2876.04 \pm 0.29$ & 223.4 \\
6 & - & 4 & \autoref{Eq:PLDampedSinusoid} & white & $-2185.80 \pm 0.29$ & 179.2 \\
7 & {\sc{ARES}} & 3 & \autoref{Eq:PLDampedSinusoid} & white & $-2074.01 \pm 0.24$ & 189.0 \\
8 & {\sc{ARES}} & 4 & \autoref{Eq:PLDampedSinusoid} & white & $-1713.37 \pm 0.24$ & 160.6 \\
9 & Gaussian & 3 & - & white & $-1168.18 \pm 0.29$ & 143.2 \\
10 & flattened Gaussian & 3 & - & white & $-212.20 \pm 0.29$ & 83.8 \\
11 & - & 5 & - & white & $-191.57 \pm 0.34$ & 95.7 \\
12 & Gaussian & 3 & \autoref{Eq:PLDampedSinusoid} & white & $-68.71 \pm 0.27$ & 70.6 \\
13 & {\sc{ARES}} & 5 & - & white & $-50.10 \pm 0.25$ & 82.5 \\
14 & - & 3 & - & \autoref{Eq:CovModelScaledRadiometerPlusWN} & $34.34 \pm 0.29$ & 252.7 \\
15 & - & 3 & \autoref{Eq:PLDampedSinusoid} & \autoref{Eq:CovModelScaledRadiometerPlusWN} & $43.35 \pm 0.24$ & 218.2 \\
16 & - & 6 & - & white & $43.80 \pm 0.32$ & 65.2 \\
17 & - & 5 & \autoref{Eq:PLDampedSinusoid} & white & $48.75 \pm 0.28$ & 71.9 \\
18 & {\sc{ARES}} & 3 & - & \autoref{Eq:CovModelScaledRadiometerPlusWN} & $64.32 \pm 0.21$ & 222.9 \\
19 & - & 4 & - & \autoref{Eq:CovModelScaledRadiometerPlusWN} & $71.80 \pm 0.28$ & 243.2 \\
20 & {\sc{ARES}} & 3 & \autoref{Eq:PLDampedSinusoid} & \autoref{Eq:CovModelScaledRadiometerPlusWN} & $72.18 \pm 0.20$ & 187.5 \\
21 & {\sc{ARES}} & 6 & - & white & $84.21 \pm 0.25$ & 60.3 \\
22 & {\sc{ARES}} & 4 & - & \autoref{Eq:CovModelScaledRadiometerPlusWN} & $84.45 \pm 0.22$ & 221.0 \\
23 & - & 4 & \autoref{Eq:PLDampedSinusoid} & \autoref{Eq:CovModelScaledRadiometerPlusWN} & $101.14 \pm 0.24$ & 193.0 \\
24 & {\sc{ARES}} & 4 & \autoref{Eq:PLDampedSinusoid} & \autoref{Eq:CovModelScaledRadiometerPlusWN} & $116.21 \pm 0.21$ & 188.1 \\
25 & {\sc{ARES}} & 5 & \autoref{Eq:PLDampedSinusoid} & white & $156.53 \pm 0.23$ & 56.9 \\
26 & Gaussian & 4 & - & white & $171.97 \pm 0.30$ & 47.2 \\
27 & - & 5 & - & \autoref{Eq:CovModelScaledRadiometerPlusWN} & $182.16 \pm 0.28$ & 95.5 \\
28 & Gaussian & 3 & - & \autoref{Eq:CovModelScaledRadiometerPlusWN} & $187.72 \pm 0.24$ & 140.0 \\
29 & Gaussian & 5 & - & white & $191.76 \pm 0.28$ & 44.1 \\
30 & Gaussian & 6 & - & white & $194.34 \pm 0.28$ & 44.0 \\
31 & {\sc{ARES}} & 5 & - & \autoref{Eq:CovModelScaledRadiometerPlusWN} & $194.62 \pm 0.22$ & 82.7 \\
32 & flattened Gaussian & 3 & - & \autoref{Eq:CovModelScaledRadiometerPlusWN} & $202.11 \pm 0.24$ & 86.8 \\
33 & - & 5 & \autoref{Eq:PLDampedSinusoid} & \autoref{Eq:CovModelScaledRadiometerPlusWN} & $205.70 \pm 0.24$ & 71.6 \\
34 & - & 7 & - & white & $213.63 \pm 0.31$ & 41.6 \\
35 & - & 8 & - & white & $214.94 \pm 0.29$ & 41.1 \\
36 & - & 9 & - & white & $216.24 \pm 0.28$ & 40.6 \\
37 & - & 10 & - & white & $217.74 \pm 0.26$ & 40.5 \\
38 & - & 6 & - & \autoref{Eq:CovModelScaledRadiometerPlusWN} & $222.50 \pm 0.27$ & 67.9 \\
39 & {\sc{ARES}} & 6 & - & \autoref{Eq:CovModelScaledRadiometerPlusWN} & $225.43 \pm 0.22$ & 63.0 \\
40 & {\sc{ARES}} & 5 & \autoref{Eq:PLDampedSinusoid} & \autoref{Eq:CovModelScaledRadiometerPlusWN} & $226.01 \pm 0.21$ & 57.0 \\
41 & {\sc{ARES}} & 8 & - & white & $233.48 \pm 0.23$ & 38.4 \\
42 & {\sc{ARES}} & 7 & - & white & $233.64 \pm 0.24$ & 38.6 \\
43 & {\sc{ARES}} & 9 & - & white & $233.94 \pm 0.22$ & 38.2 \\
44 & {\sc{ARES}} & 10 & - & white & $234.22 \pm 0.22$ & 38.1 \\
45 & Gaussian & 3 & \autoref{Eq:PLDampedSinusoid} & \autoref{Eq:CovModelScaledRadiometerPlusWN} & $234.96 \pm 0.23$ & 88.5 \\
46 & Gaussian & 6 & - & \autoref{Eq:CovModelScaledRadiometerPlusWN} & $243.08 \pm 0.25$ & 43.9 \\
47 & Gaussian & 4 & - & \autoref{Eq:CovModelScaledRadiometerPlusWN} & $243.92 \pm 0.26$ & 49.4 \\
48 & Gaussian & 5 & - & \autoref{Eq:CovModelScaledRadiometerPlusWN} & $246.36 \pm 0.25$ & 44.3 \\
49 & - & 9 & - & \autoref{Eq:CovModelScaledRadiometerPlusWN} & $249.97 \pm 0.24$ & 40.9 \\
50 & - & 8 & - & \autoref{Eq:CovModelScaledRadiometerPlusWN} & $250.01 \pm 0.26$ & 41.1 \\
51 & - & 10 & - & \autoref{Eq:CovModelScaledRadiometerPlusWN} & $250.43 \pm 0.23$ & 40.4 \\
52 & - & 7 & - & \autoref{Eq:CovModelScaledRadiometerPlusWN} & $250.68 \pm 0.27$ & 41.6 \\
53 & {\sc{ARES}} & 8 & - & \autoref{Eq:CovModelScaledRadiometerPlusWN} & $255.42 \pm 0.21$ & 38.4 \\
54 & {\sc{ARES}} & 9 & - & \autoref{Eq:CovModelScaledRadiometerPlusWN} & $255.88 \pm 0.21$ & 38.6 \\
55 & {\sc{ARES}} & 10 & - & \autoref{Eq:CovModelScaledRadiometerPlusWN} & $255.89 \pm 0.20$ & 38.4 \\
\bottomrule
\end{tabular}
}
\label{Tab:ModelEvidences1}
\end{table*}

\begin{table*}
\caption{\autoref{Tab:ModelEvidences1} continued. }.
\centerline{
\begin{tabular}{l l l l l l l }
\toprule
Model number & Global signal & Log-polynomial & Damped sinusoid, & Noise model, & log(evidence) & Residual RMS             \\
           & model, $\sdelta T_\mathrm{b}$         & order, $\bar{T}_\mathrm{Fg}$          &  $T_\mathrm{cal}$               & N          &               & [mK]                   \\
\midrule
56 & {\sc{ARES}} & 7 & - & \autoref{Eq:CovModelScaledRadiometerPlusWN} & $256.87 \pm 0.22$ & 38.6 \\
57 & flattened Gaussian & 4 & - & white & $275.35 \pm 0.29$ & 38.7 \\
58 & flattened Gaussian & 4 & - & \autoref{Eq:CovModelScaledRadiometerPlusWN} & $277.32 \pm 0.26$ & 38.9 \\
59 & - & 6 & \autoref{Eq:PLDampedSinusoid} & \autoref{Eq:CovModelScaledRadiometerPlusWN} & $280.69 \pm 0.25$ & 33.6 \\
60 & - & 6 & \autoref{Eq:PLDampedSinusoid} & white & $281.27 \pm 0.27$ & 33.6 \\
61 & flattened Gaussian & 3 & \autoref{Eq:PLDampedSinusoid} & white & $292.42 \pm 0.25$ & 29.9 \\
62 & {\sc{ARES}} & 6 & \autoref{Eq:PLDampedSinusoid} & \autoref{Eq:CovModelScaledRadiometerPlusWN} & $294.27 \pm 0.22$ & 28.6 \\
63 & Gaussian & 7 & - & white & $296.02 \pm 0.29$ & 27.0 \\
64 & {\sc{ARES}} & 6 & \autoref{Eq:PLDampedSinusoid} & white & $296.45 \pm 0.23$ & 28.5 \\
65 & Gaussian & 7 & - & \autoref{Eq:CovModelScaledRadiometerPlusWN} & $296.82 \pm 0.26$ & 27.1 \\
66 & Gaussian & 9 & - & \autoref{Eq:CovModelScaledRadiometerPlusWN} & $298.20 \pm 0.25$ & 27.1 \\
67 & Gaussian & 8 & - & \autoref{Eq:CovModelScaledRadiometerPlusWN} & $298.63 \pm 0.25$ & 26.6 \\
68 & flattened Gaussian & 3 & \autoref{Eq:PLDampedSinusoid} & \autoref{Eq:CovModelScaledRadiometerPlusWN} & $300.67 \pm 0.23$ & 30.0 \\
69 & Gaussian & 4 & \autoref{Eq:PLDampedSinusoid} & white & $301.31 \pm 0.26$ & 28.8 \\
70 & Gaussian & 8 & - & white & $301.48 \pm 0.27$ & 26.7 \\
71 & Gaussian & 10 & - & \autoref{Eq:CovModelScaledRadiometerPlusWN} & $301.66 \pm 0.23$ & 26.9 \\
72 & Gaussian & 10 & - & white & $303.63 \pm 0.25$ & 27.0 \\
73 & Gaussian & 9 & - & white & $303.98 \pm 0.26$ & 27.1 \\
74 & Gaussian & 4 & \autoref{Eq:PLDampedSinusoid} & \autoref{Eq:CovModelScaledRadiometerPlusWN} & $309.66 \pm 0.24$ & 29.5 \\
75 & flattened Gaussian & 7 & - & white & $313.16 \pm 0.27$ & 23.6 \\
76 & flattened Gaussian & 6 & - & white & $315.16 \pm 0.28$ & 23.5 \\
77 & flattened Gaussian & 8 & - & white & $317.17 \pm 0.26$ & 22.2 \\
78 & flattened Gaussian & 10 & - & white & $317.58 \pm 0.24$ & 22.1 \\
79 & flattened Gaussian & 9 & - & white & $318.23 \pm 0.25$ & 22.0 \\
80 & flattened Gaussian & 5 & - & white & $319.22 \pm 0.28$ & 23.8 \\
81 & flattened Gaussian & 6 & \autoref{Eq:PLDampedSinusoid} & white & $319.24 \pm 0.24$ & 20.8 \\
82 & flattened Gaussian & 7 & - & \autoref{Eq:CovModelScaledRadiometerPlusWN} & $320.04 \pm 0.25$ & 23.8 \\
83 & flattened Gaussian & 9 & \autoref{Eq:PLDampedSinusoid} & white & $321.38 \pm 0.22$ & 19.8 \\
84 & flattened Gaussian & 10 & \autoref{Eq:PLDampedSinusoid} & white & $321.54 \pm 0.21$ & 20.7 \\
85 & flattened Gaussian & 8 & \autoref{Eq:PLDampedSinusoid} & white & $321.55 \pm 0.22$ & 19.9 \\
86 & flattened Gaussian & 6 & - & \autoref{Eq:CovModelScaledRadiometerPlusWN} & $321.60 \pm 0.26$ & 23.9 \\
87 & Gaussian & 6 & \autoref{Eq:PLDampedSinusoid} & white & $321.78 \pm 0.24$ & 21.0 \\
88 & flattened Gaussian & 7 & \autoref{Eq:PLDampedSinusoid} & white & $322.12 \pm 0.23$ & 20.6 \\
89 & flattened Gaussian & 4 & \autoref{Eq:PLDampedSinusoid} & white & $322.35 \pm 0.25$ & 21.1 \\
90 & flattened Gaussian & 5 & \autoref{Eq:PLDampedSinusoid} & white & $322.92 \pm 0.24$ & 20.8 \\
91 & Gaussian & 9 & \autoref{Eq:PLDampedSinusoid} & white & $323.04 \pm 0.22$ & 20.9 \\
92 & Gaussian & 10 & \autoref{Eq:PLDampedSinusoid} & white & $323.38 \pm 0.21$ & 20.8 \\
93 & - & 8 & \autoref{Eq:PLDampedSinusoid} & white & $323.42 \pm 0.25$ & 20.8 \\
94 & Gaussian & 8 & \autoref{Eq:PLDampedSinusoid} & white & $323.59 \pm 0.23$ & 20.9 \\
95 & Gaussian & 7 & \autoref{Eq:PLDampedSinusoid} & white & $323.60 \pm 0.23$ & 20.7 \\
96 & {\sc{ARES}} & 9 & \autoref{Eq:PLDampedSinusoid} & white & $323.83 \pm 0.21$ & 20.9 \\
97 & {\sc{ARES}} & 10 & \autoref{Eq:PLDampedSinusoid} & white & $324.01 \pm 0.20$ & 21.0 \\
98 & - & 10 & \autoref{Eq:PLDampedSinusoid} & white & $324.05 \pm 0.23$ & 21.1 \\
99 & {\sc{ARES}} & 8 & \autoref{Eq:PLDampedSinusoid} & white & $324.07 \pm 0.21$ & 20.9 \\
100 & - & 9 & \autoref{Eq:PLDampedSinusoid} & white & $324.15 \pm 0.24$ & 20.8 \\
101 & {\sc{ARES}} & 7 & \autoref{Eq:PLDampedSinusoid} & white & $324.25 \pm 0.22$ & 20.7 \\
102 & - & 7 & \autoref{Eq:PLDampedSinusoid} & white & $324.62 \pm 0.26$ & 20.7 \\
103 & Gaussian & 5 & \autoref{Eq:PLDampedSinusoid} & white & $324.64 \pm 0.25$ & 21.0 \\
104 & flattened Gaussian & 8 & - & \autoref{Eq:CovModelScaledRadiometerPlusWN} & $324.94 \pm 0.25$ & 22.0 \\
105 & flattened Gaussian & 10 & - & \autoref{Eq:CovModelScaledRadiometerPlusWN} & $325.13 \pm 0.23$ & 21.9 \\
106 & flattened Gaussian & 9 & - & \autoref{Eq:CovModelScaledRadiometerPlusWN} & $325.18 \pm 0.24$ & 21.8 \\
107 & flattened Gaussian & 5 & - & \autoref{Eq:CovModelScaledRadiometerPlusWN} & $326.89 \pm 0.27$ & 23.6 \\
108 & Gaussian & 9 & \autoref{Eq:PLDampedSinusoid} & \autoref{Eq:CovModelScaledRadiometerPlusWN} & $330.71 \pm 0.22$ & 20.6 \\
109 & flattened Gaussian & 4 & \autoref{Eq:PLDampedSinusoid} & \autoref{Eq:CovModelScaledRadiometerPlusWN} & $330.75 \pm 0.25$ & 21.3 \\
110 & Gaussian & 10 & \autoref{Eq:PLDampedSinusoid} & \autoref{Eq:CovModelScaledRadiometerPlusWN} & $330.94 \pm 0.21$ & 20.8 \\
\bottomrule
\end{tabular}
}
\label{Tab:ModelEvidences2}
\end{table*}

\begin{table*}
\caption{\autoref{Tab:ModelEvidences1} continued. }.
\centerline{
\begin{tabular}{l l l l l l l }
\toprule
Model number & Global signal & Log-polynomial & Damped sinusoid, & Noise model, & log(evidence) & Residual RMS             \\
           & model, $\sdelta T_\mathrm{b}$         & order, $\bar{T}_\mathrm{Fg}$          &  $T_\mathrm{cal}$               & N          &               & [mK]                   \\
\midrule
111 & flattened Gaussian & 6 & \autoref{Eq:PLDampedSinusoid} & \autoref{Eq:CovModelScaledRadiometerPlusWN} & $332.17 \pm 0.24$ & 20.4 \\
112 & Gaussian & 6 & \autoref{Eq:PLDampedSinusoid} & \autoref{Eq:CovModelScaledRadiometerPlusWN} & $332.21 \pm 0.24$ & 21.0 \\
113 & flattened Gaussian & 8 & \autoref{Eq:PLDampedSinusoid} & \autoref{Eq:CovModelScaledRadiometerPlusWN} & $332.37 \pm 0.22$ & 19.7 \\
114 & flattened Gaussian & 5 & \autoref{Eq:PLDampedSinusoid} & \autoref{Eq:CovModelScaledRadiometerPlusWN} & $332.63 \pm 0.24$ & 20.4 \\
115 & Gaussian & 8 & \autoref{Eq:PLDampedSinusoid} & \autoref{Eq:CovModelScaledRadiometerPlusWN} & $332.81 \pm 0.23$ & 20.7 \\
116 & flattened Gaussian & 9 & \autoref{Eq:PLDampedSinusoid} & \autoref{Eq:CovModelScaledRadiometerPlusWN} & $333.58 \pm 0.22$ & 19.6 \\
117 & {\sc{ARES}} & 10 & \autoref{Eq:PLDampedSinusoid} & \autoref{Eq:CovModelScaledRadiometerPlusWN} & $334.07 \pm 0.20$ & 20.7 \\
118 & - & 9 & \autoref{Eq:PLDampedSinusoid} & \autoref{Eq:CovModelScaledRadiometerPlusWN} & $334.08 \pm 0.24$ & 20.8 \\
119 & - & 10 & \autoref{Eq:PLDampedSinusoid} & \autoref{Eq:CovModelScaledRadiometerPlusWN} & $334.08 \pm 0.23$ & 20.7 \\
120 & Gaussian & 5 & \autoref{Eq:PLDampedSinusoid} & \autoref{Eq:CovModelScaledRadiometerPlusWN} & $334.18 \pm 0.24$ & 21.1 \\
121 & {\sc{ARES}} & 9 & \autoref{Eq:PLDampedSinusoid} & \autoref{Eq:CovModelScaledRadiometerPlusWN} & $334.25 \pm 0.21$ & 20.8 \\
122 & {\sc{ARES}} & 7 & \autoref{Eq:PLDampedSinusoid} & \autoref{Eq:CovModelScaledRadiometerPlusWN} & $334.28 \pm 0.22$ & 20.9 \\
123 & {\sc{ARES}} & 8 & \autoref{Eq:PLDampedSinusoid} & \autoref{Eq:CovModelScaledRadiometerPlusWN} & $334.40 \pm 0.21$ & 20.8 \\
124 & - & 8 & \autoref{Eq:PLDampedSinusoid} & \autoref{Eq:CovModelScaledRadiometerPlusWN} & $334.48 \pm 0.25$ & 20.7 \\
125 & - & 7 & \autoref{Eq:PLDampedSinusoid} & \autoref{Eq:CovModelScaledRadiometerPlusWN} & $334.64 \pm 0.26$ & 20.9 \\
126 & flattened Gaussian & 10 & \autoref{Eq:PLDampedSinusoid} & \autoref{Eq:CovModelScaledRadiometerPlusWN} & $334.97 \pm 0.22$ & 19.7 \\
127 & Gaussian & 7 & \autoref{Eq:PLDampedSinusoid} & \autoref{Eq:CovModelScaledRadiometerPlusWN} & $335.09 \pm 0.23$ & 20.9 \\
128 & flattened Gaussian & 7 & \autoref{Eq:PLDampedSinusoid} & \autoref{Eq:CovModelScaledRadiometerPlusWN} & $336.17 \pm 0.23$ & 19.8 \\
\bottomrule
\end{tabular}
}
\label{Tab:ModelEvidences2}
\end{table*}

In \autoref{Tab:ModelPriors}, we list the full set of free parameters that we sample from in the different models used in the analysis, along with the minimum and maximum values of the priors on those parameters. Throughout the analysis presented in this section, we assume uniform prior probability distributions on the model parameters. In all models, we fit for the model components simultaneously, sample from the parameters and calculate the Bayesian evidence of the model using nested sampling as implemented by the {\sc{PolyChord}} algorithm \citep{2015MNRAS.453.4384H, 2015MNRAS.450L..61H}.

In the analysis presented in this section, we compare models comprised of one of each of the following components:
\begin{enumerate}
\item \textit{$\sdelta T_\mathrm{b}$ (4 models)} - no detectable global signal, or one of the three global signal parametrisations described in \autoref{EDGESDataModel}: a flattened Gaussian (FG), Gaussian (G) or {\sc{ARES}} simulation.
\item $\bar{T}_\mathrm{Fg}$ \textit{(8 models)} - log-polynomial models  between 3rd and 10th order. Here, we use a 3rd order log-polynomial as a minimum complexity model of the intrinsic foregrounds in the scenario that there are negligible calibration errors, or modulation by ionospheric effects. We use higher order log-polynomials as a combined model for  intrinsic foreground spectral structure modulated by the ionosphere and contamination by certain classes of calibration systematics, as detailed in \autoref{EDGESData}.
\item $T_\mathrm{cal}$ \textit{(2 models)} - we consider models both with and without the inclusion of a power law damped sinusoidal systematic described by \autoref{Eq:PLDampedSinusoid}.
\item $\mathbfss{N}$ \textit{(2 models)} - \autoref{Eq:CovModelScaledRadiometerPlusWN} describes our primary model for the covariance structure of the noise in the EDGES low-band data. We compare this to an intrinsic white noise model\footnote{We incorporate the effects of the non-uniform channel weights (the fraction of data integrated in each spectral bin) of the data used in the analysis for both noise models.}, as used in the analyses of the EDGES data carried out by B18, H18 and S19.
\end{enumerate}

In total, this yields 128 models. The derived Bayesian evidences and RMS of the $(\mathbfit{d} - \mathbfit{m}(\sTheta_\mathrm{MAP}))$ residuals, with $\sTheta_\mathrm{MAP}$ the maximum a posteriori parameters of the model for the full set of models considered in the analysis, are listed in \autoref{Tab:ModelEvidences1}.

Out of the total set of models analysed, the model with the highest Bayesian evidence (m128) includes a flattened Gaussian parametrisation for the global 21 cm signal, a 7th order log-polynomial $\bar{T}_\mathrm{Fg}$ model, a power law damped sinusoid and a generalised noise model. Additionally, there are a further twelve models (m116-m127) with Bayesian evidence values sufficiently high to not be strongly disfavoured relative to m128 ($\log(B)>3$). All of these models have a power law damped sinusoidal component and a generalised noise model. Furthermore, of the total set of models analysed, those including a power law damped sinusoidal component and a generalised noise model are decisively preferred (e.g. \citealt{KandR}): the highest evidence models that do not include a power law damped sinusoidal (m107) or a generalised noise model (m103) have $\log(B) \sim 9.3$ and $11.5$ relative to the highest evidence model, respectively.

This preference for models including higher order log-polynomial components, a damped sinusoidal component and a noise level significantly in excess of that which is theoretically expected, is indicative of spectral structure in the EDGES low-band data that is not consistent with a combination of intrinsic foreground emission (or even foregrounds modulated by the ionosphere), an imperfectly calibrated-but-smooth-spectrum receiver gain and a global 21 cm signal. Instead, our model comparison demonstrates that our model components, whach are designed to describe a calibration systematic error dominated by an error in the amplitude, phase or period of the sinusoidal undulation component of the electromagnetic beam model for the gain at a particular beam angle and lower level contributions from a range of other beam angles, an imperfect beam chromaticity correction to the data and potentially polarised foreground emission describe statistically important features of the data.

Of the 13 highest evidence models, there are 3 with flattened Gaussian and 2 Gaussian global 21 cm signal parametrisations, 4 have no global signal component and 4 use global 21 cm signals derived from our atlas of 100,000 {\sc{ARES}} simulations. This shows that neither a specific shape of the global 21 signal, nor, indeed, its detection, are preferred by a statistically significant margin in the publicly available EDGES low-band data set. Instead, the preference for including calibration systematic based models, in the form of an explicit power law damped sinusoidal feature, high-order (greater than 5th order) log-polynomials and a generalised noise model, is of greater statistical significance.

Nevertheless, examination of the properties of the signal components in aggregate, or within the set of high evidence models, can reveal physical trends of interest. We, therefore, proceed with such an examination in the following sections.

\section{Analysis of preferred models}
\label{AnalysisOfPreferredModels}

\subsection{Global signal types}
\label{GST}

\begin{table*}
\caption{MAP parameter values of the highest evidence model of each of the four global signal model scenarios (flattened Gaussian, m128; Gaussian, m127; {\sc{ARES}}, m123; no global signal model, m125) and the model components the parameters are associated with.}.
\centerline{
\begin{tabular}{l l l l l l l}
\toprule
Model component & Parameter & Unit & m128 MAP parameter  & m127 MAP parameter  & m125 MAP parameter  & m123 MAP parameter    \\
                &           &      & estimate            & estimate            & estimate            & estimate              \\
$\bar{T}_\mathrm{Fg}$                 &  $d_{0}$  & $\log_{10}(T[\mathrm{K}])$ &  $3.67$     &  $3.67$     &  $3.67$    &  $3.67$    \\
                                      &  $d_{1}$  & - &  $-2.55$    &  $-2.55$    &  $-2.55$    &  $-2.55$   \\
                                      &  $d_{2}$  & - &  $0.02$     &  $0.02$     &  $0.03$     &  $0.02$    \\
                                      &  $d_{3}$  & - &  $-0.40$    &  $-0.29$    &  $-0.59$    &  $-0.22$   \\
                                      &  $d_{4}$  & - &  $-2.13$    &  $-3.91$    &  $0.23$     &  $-4.49$   \\
                                      &  $d_{5}$  & - &  $16.27$    &  $24.45$    &  $-1.10$    &  $26.75$   \\
                                      &  $d_{6}$  & - &  $-25.53$   &  $-37.49$   &  $35.76$    &  $-40.73$  \\
                                      &  $d_{7}$  & - &  -          &  -          &  $-79.12$   &  -  \\
\midrule
$\sdelta T_\mathrm{b}$: FG            &  $\tau$   & - &  $25.3$   &  -        &  -   &  -       \\
$\sdelta T_\mathrm{b}$: FG, G         &  $A$      & $\mathrm{K}$ &  $0.16$   &  $0.14$   &  -   &  -  \\
                                      &  $\nu_0$  & $\mathrm{MHz}$ &  $78.8$   &  $74.4$   &  -   &  -  \\
                                      &  $w$      & $\mathrm{MHz}$ &  $17.9$   &  $32.3$   &  -   &  -  \\
\midrule
$\sdelta T_\mathrm{b}$: {\sc{ARES}}          &  $f_\mathrm{X}$    & - &  -  &  -  &  $25.4$  &  -  \\
                                      &  $f_{\star}$       & - &  -  &  -  &  $0.75$  &  -  \\
                                      &  $T_\mathrm{min}$  & $\mathrm{K}$ &  -  &  -  &  $9.7\times10^{4}$  &  - \\
                                      &  $N_\mathrm{lw}$   & - &  -  &  -  &  $3.2\times10^{4}$  &  - \\
                                      &  $N_\mathrm{ion}$  & - &  -  &  -  &  $1.2\times10^{5}$  &  - \\
\midrule
$T_\mathrm{cal}$                      &  $\bar{a}_{0}$  & $\log_{10}(T[\mathrm{K}])$ &  $-1.45$  &  $-1.23$  &  $-1.23$  &  $-1.23$ \\
                                      &  $\bar{a}_{1}$  & $\log_{10}(T[\mathrm{K}])$ &  $-8.49$  &  $-5.13$  &  $-9.84$  &  $-4.49$ \\
                                      &  $P$            & $\mathrm{MHz}$ &  $12.44$  &  $12.50$  &  $12.49$  &  $12.50$ \\
                                      &  $b$            & - &  $-2.63$  &  $-1.02$  &  $-1.05$  &  $-1.18$ \\
\midrule
$N$                                   &  $\alpha_\mathrm{rn}$  & - &  $2.26$   &  $2.33$   &  $2.25$   &  $2.20$ \\
                                      &  $\alpha_\mathrm{wn}$  & $\mathrm{K}$ &  $0.010$  &  $0.012$  &  $0.012$  &  $0.010$ \\
\bottomrule
\end{tabular}
}
\label{Tab:MAPParams}
\end{table*}

\begin{figure*}
	\centerline{
	\includegraphics[width=0.5\textwidth]{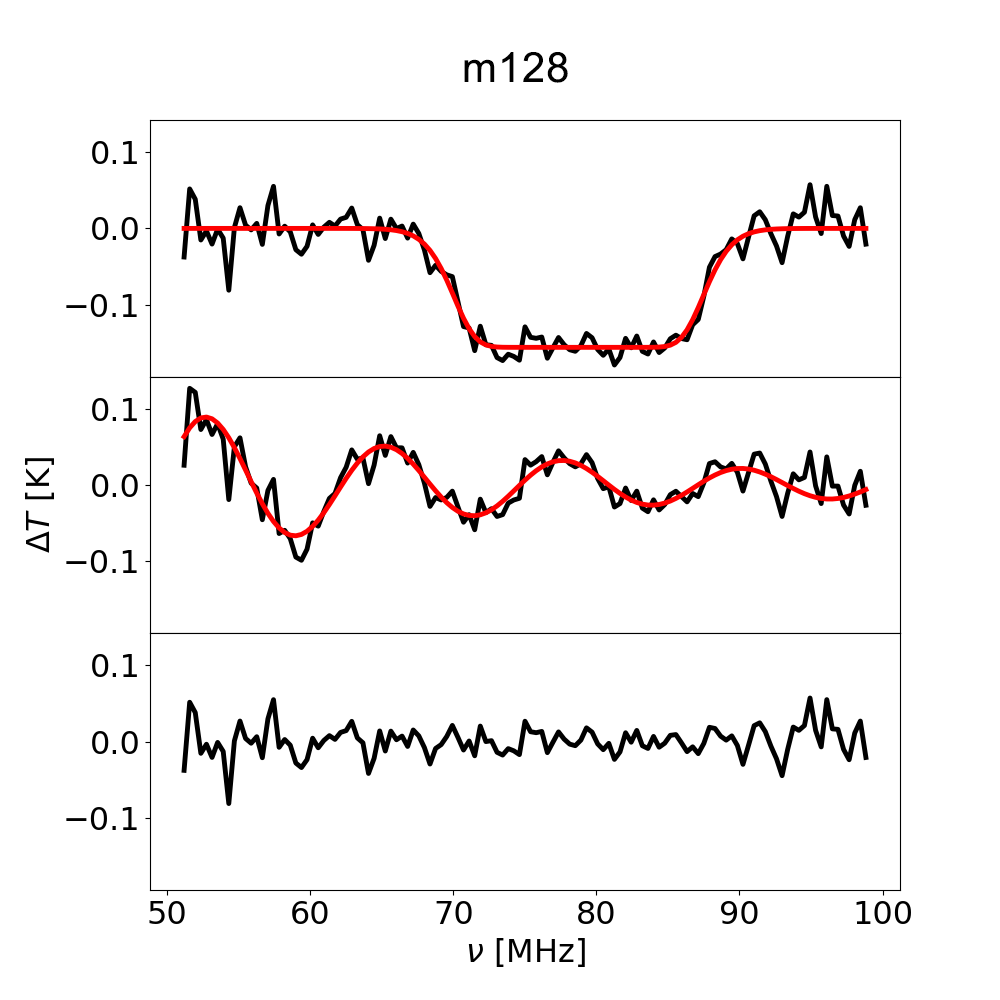}
	\includegraphics[width=0.5\textwidth]{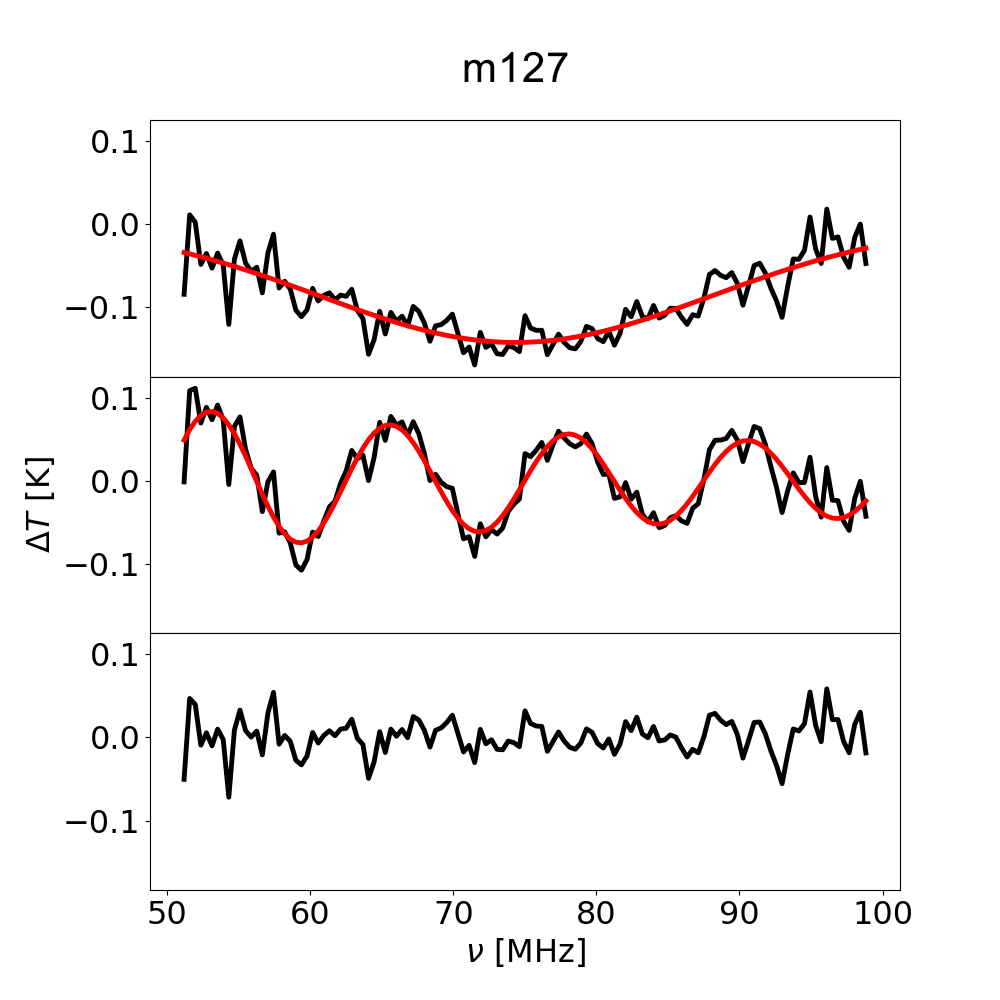}
	}
	\centerline{
	\includegraphics[width=0.5\textwidth]{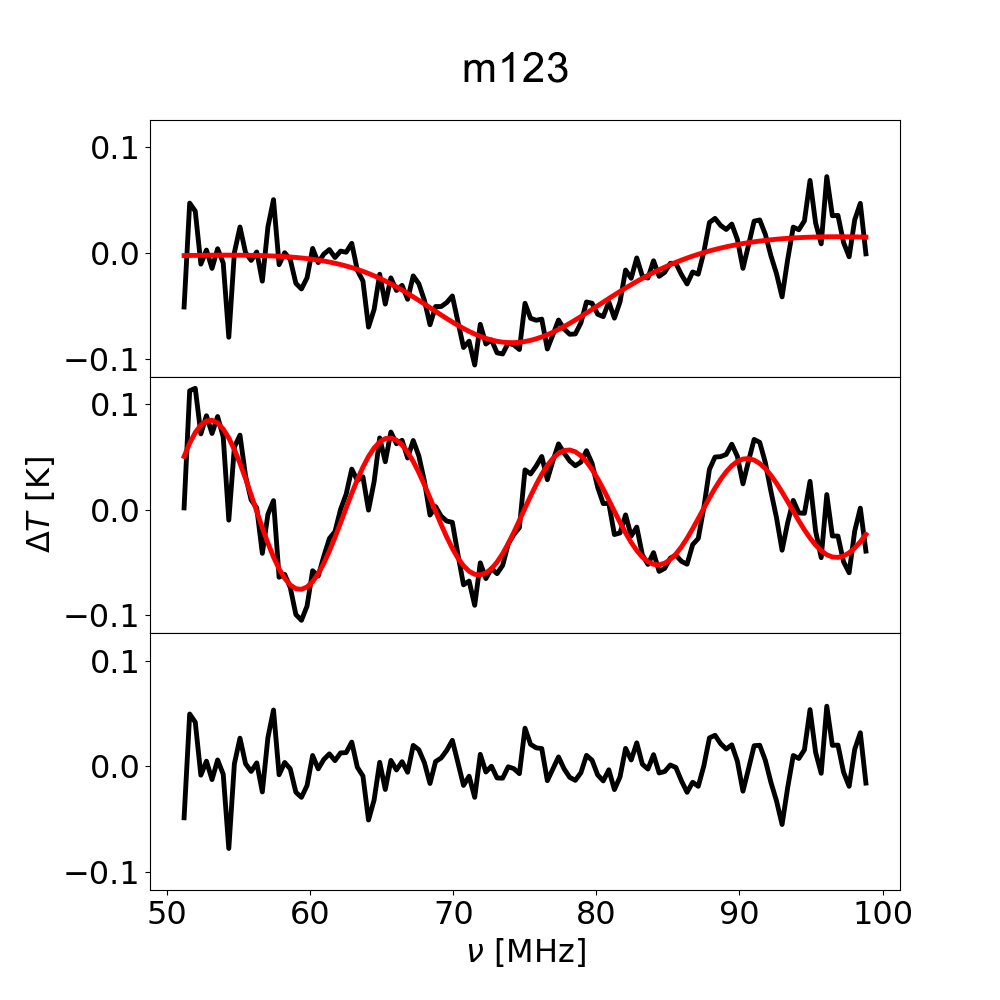}
	\includegraphics[width=0.5\textwidth]{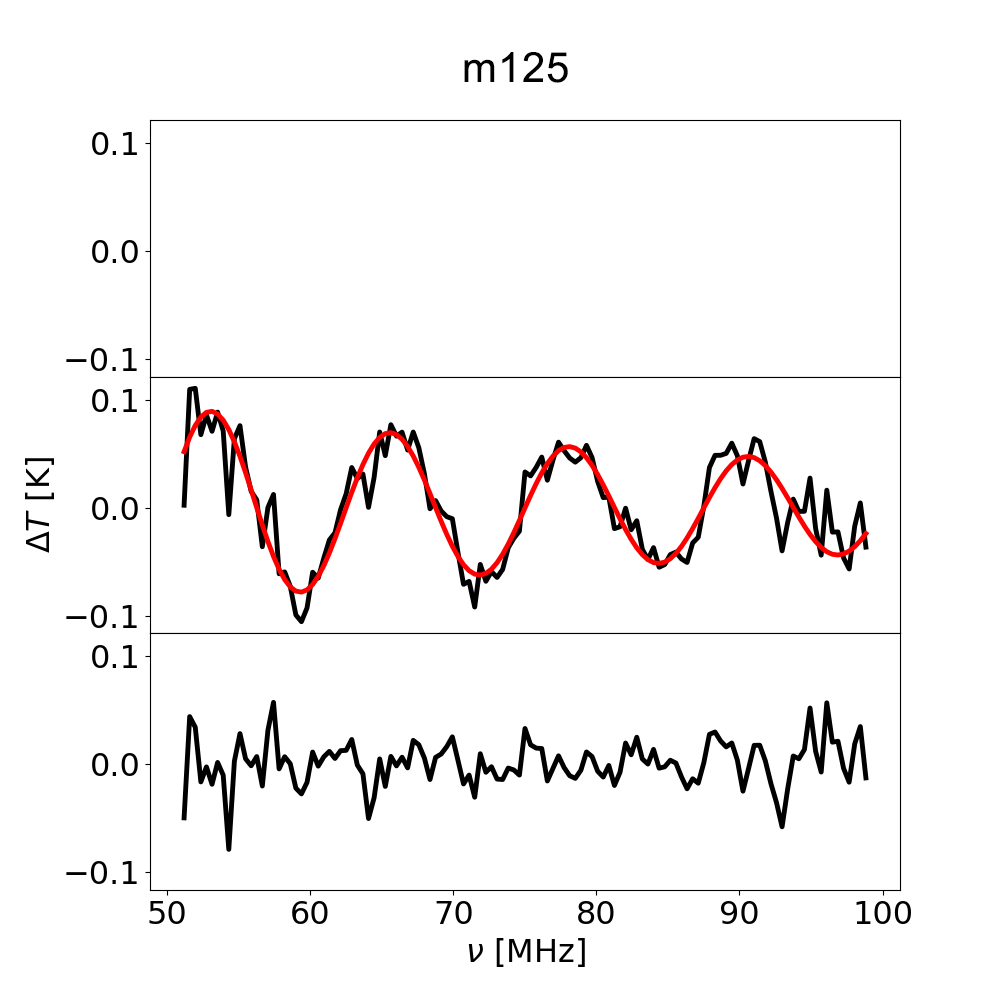}
	}
\caption{MAP signal components associated with the highest evidence models of the EDGES low-band data set in each of the four global 21 cm signal model scenarios (clockwise from top left: flattened Gaussian, m128; Gaussian, m127; no global 21 cm signal model, m125; {\sc{ARES}}, m123). Each subplot displays the MAP global 21 cm signal (top) and power law damped systematic (center) model components, in red. The corresponding $(d-T_\mathrm{Fg}-T_\mathrm{sys})$ (top) and $(d-T_\mathrm{Fg}-T_\mathrm{21})$ (center) data residuals are shown in black. The $(d-T_\mathrm{Fg}-T_\mathrm{sys}-T_{21})$ residuals of the full MAP models are shown in the bottom panels. The top panel is blank in the bottom left plot due to the absence of a global 21 cm signal component in m123. Clockwise from top left, the RMS of the residuals are 19.8, 20.8, 20.8 and 20.9 mK, respectively.}
\label{ForegroundResiduals}
\end{figure*}

Above, we identify thirteen preferred models for the EDGES low-band data, which all provide comparably good descriptions to the highest evidence model (m116-m128; those with $\log(B)<3$ relative to m128, the highest evidence model in our analysis). In this section, we look at four of these models in more detail. Specifically, we focus on the highest evidence model of each of the four global signal model scenarios: flattened Gaussian, m128; Gaussian, m127; {\sc{ARES}}, m123 and no global signal, m125. In \autoref{Tab:MAPParams}, we list the maximum a posteriori (MAP) parameter values of the models and the model components the parameters are associated with. In \autoref{ForegroundResiduals}, we plot the MAP global 21 cm signal, power law damped systematic model components and data residuals for each of the models. 

Comparing the models, it is evident that a broad range of MAP global 21 cm signal components, with significant differences both in shape and width, can provide a comparably good fit to the data when jointly estimated with the other signal components necessary to produce high Bayesian evidence models of the spectral structure of the EDGES low-band data. These include:
\begin{itemize}
\item \textit{In model m128 -} a highly flattened Gaussian ($\tau_\mathrm{MAP} = 25.3$) centered on $\nu_{0}=79~\mathrm{MHz}$ ($z=17$). A physical interpretation of this model would imply very rapid Wouthuysen--Field coupling of $T_{S}$ to $T_{K}$ at $z\sim19$, followed by a period of approximately $75~\mathrm{Myr}$ in which there was minimal input of heating of the IGM by sources (with heat input proportional to $(1+z)$ and, thus, sufficient only to slow the adiabatic cooling of the gas and maintain a constant differential brightness relative to the cooling background radiation temperature\footnote{This assumes that the background radiation temperature is CMB dominated during this period and thus drops as $(1+z)$.}) and ending at $z\sim15$, when the gas was rapidly heated, nullifying the absorption feature.
\item  \textit{In model m126 -} an extremely broad Gaussian feature ($w=32.3~\mathrm{MHz}$) centered on $\nu_{0}=74~\mathrm{MHz}$ ($z=18$). A physical interpretation of the model implies a slow Wouthuysen--Field coupling of $T_{S}$ to $T_{K}$, beginning as early as $z\sim28$, and accompanied by a gradual heating of the IGM by sources over an extended period of $\sim200~\mathrm{Myr}$, with $T_{S}$ and $T_{R}$ reaching equality at $z=13$.
\item  \textit{In model m123 -} a physically motivated absorption signature sourced by galaxies with physical parameters listed in \autoref{Tab:MAPParams}. The signal is centered on $\nu_{0}=74~\mathrm{MHz}$ ($z=18$) and begins at $\nu\sim62~\mathrm{MHz}$ ($z=22$). In this case, heating of the IGM by sources raises $T_{S}$ to the level of $T_{R}$ by $z=15$ and continues to raise the temperature, producing a signal in emission at $z<15$.
\end{itemize}

Despite the differences in the shapes of the MAP global 21 cm signals, there are also a number of important commonalities between the components of the different models:
\begin{enumerate}
\item in each case, the MAP noise model comprises of a radiometer noise component with an amplitude 2.2--2.3 times that expected for the intrinsic radiometer noise level (see \autoref{CovarianceMatrixModel}), plus a 10--12 mK white noise level.
\item a power law damped sine wave feature with a period of $\sim12.5~\mathrm{MHz}$ is present in all cases. In m127, m125 and m123, the features have a consistent amplitude of $\sim60~\mathrm{mK}$ at the reference frequency, $\nu_{0}=75~\mathrm{MHz}$, and a power law damping index of $\sim-1.1$, compared to a feature with an amplitude of $\sim35~\mathrm{mK}$ at the reference frequency, $\nu_{0}=75~\mathrm{MHz}$, and a power law damping index of $\sim-2.6$ in m128. It can be seen in \autoref{ForegroundResiduals} that this results in sinusoidal features with similar amplitudes in all four cases at the low frequency end of the band, but a more significant contraction in the amplitude of the systematic feature (which must therefore be absorbed by the foreground and global signal model components) at higher frequencies in m128.
\item The low order polynomial coefficients, up to 3rd order, which are expected to be most important for describing the intrinsic spectral structure of the foregrounds, are almost identical in all four cases, with a marginally higher running of the spectral index in m125. In contrast, greater divergence is evident in the higher order terms, which we expect to be associated with imperfectly calibrated antenna chromaticity and the possible presence of polarised foreground emission. The relatively broad prior range on these parameters (see \autoref{Tab:ModelPriors}) is necessary to accurately represent the spectral structure in the EDGES low-band data; however, it is also the source of comparably good fits being obtainable with different shape global 21 cm signal models. To better constrain the shape of the global 21 cm signal, it would be necessary to place more stringent priors on these parameters. 
 
The most direct way to remove this covariance would be through eliminating residual chromatic calibration artifacts during data calibration, thus enabling the prior range on these parameters to be restricted to small values or, potentially, removing the need for the higher order terms in the log-polynomial all together. However, the accuracy of the sky model and electromagnetic beam model required to achieve this will be a function of the specific spectral structure of the EDGES beam and may be challenging (e.g. see \autoref{EDGESBeamChromaticityCorrectionErrors}) If it is the case that the contaminating chromatic structure derives from a limited number of well defined systematic structures, and these could be better determined, it is possible this could be achieved by jointly fitting for these structures in the analysis, in combination with a simpler log-polynomial model purely of the intrinsic foregrounds. As mentioned in \autoref{CalibrationSystematicModel}, ideally, solving for the required calibration model would be carried out simultaneously with estimation of the signal components, so that the multiplicative effect of the gain model and its uncertainties can be correctly accounted for.
\item In the three models that include a global 21 cm signal, the MAP amplitudes of the signal are consistent with the standard cosmological model ($0<A<209~\mathrm{mK}$) in each case (c.f. the {\sc{ARES}} models have amplitudes that are consistent with the standard cosmological model by construction). We explore this further in the next section.
\end{enumerate}

\subsection{Global signal parameters}
\label{EP}

\begin{figure*}
	\centerline{
	\includegraphics[width=0.5\textwidth]{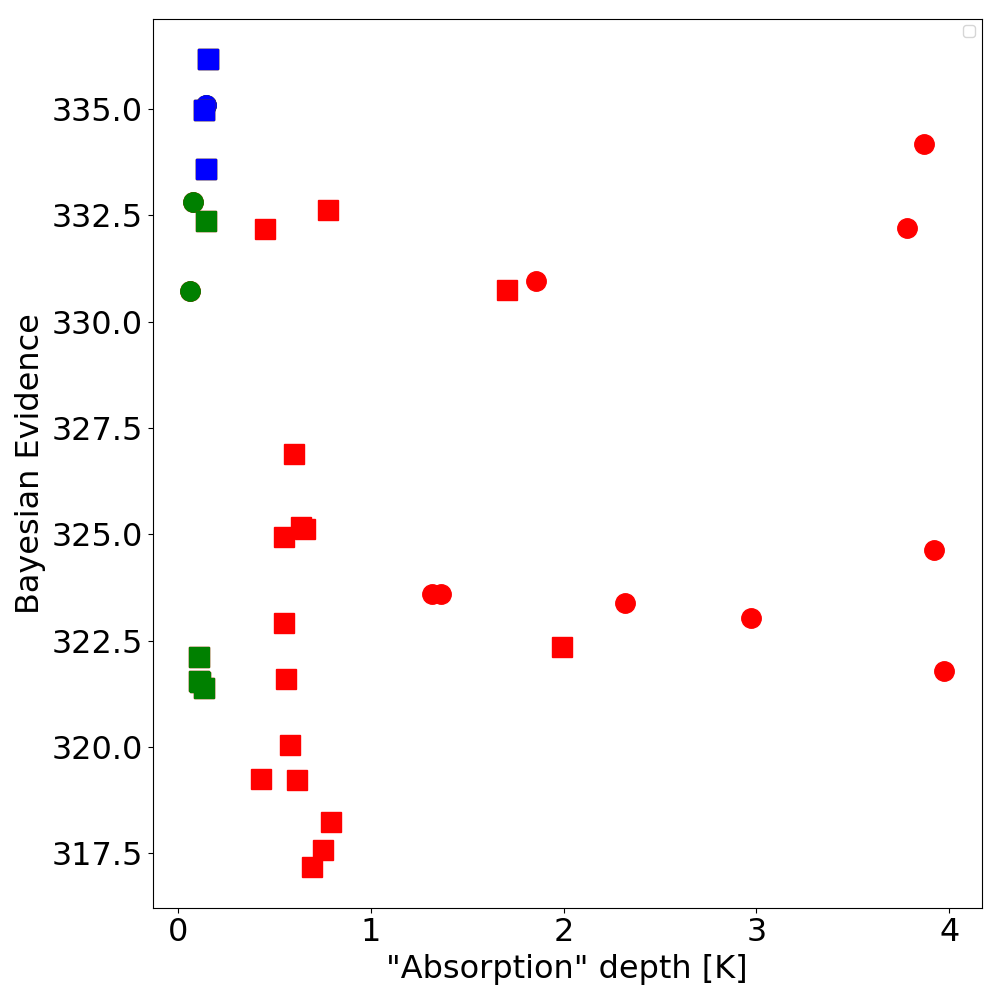}
	\includegraphics[width=0.5\textwidth]{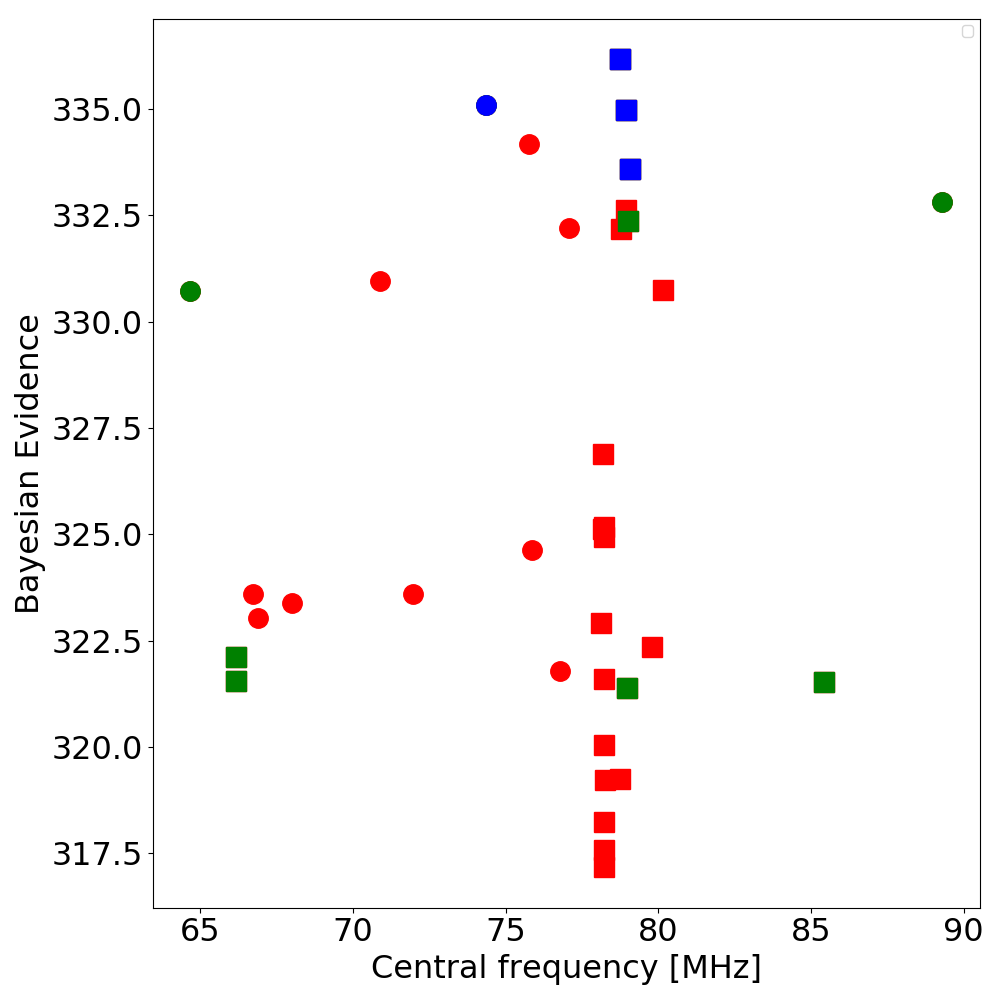}}
	\centerline{
	\includegraphics[width=0.5\textwidth]{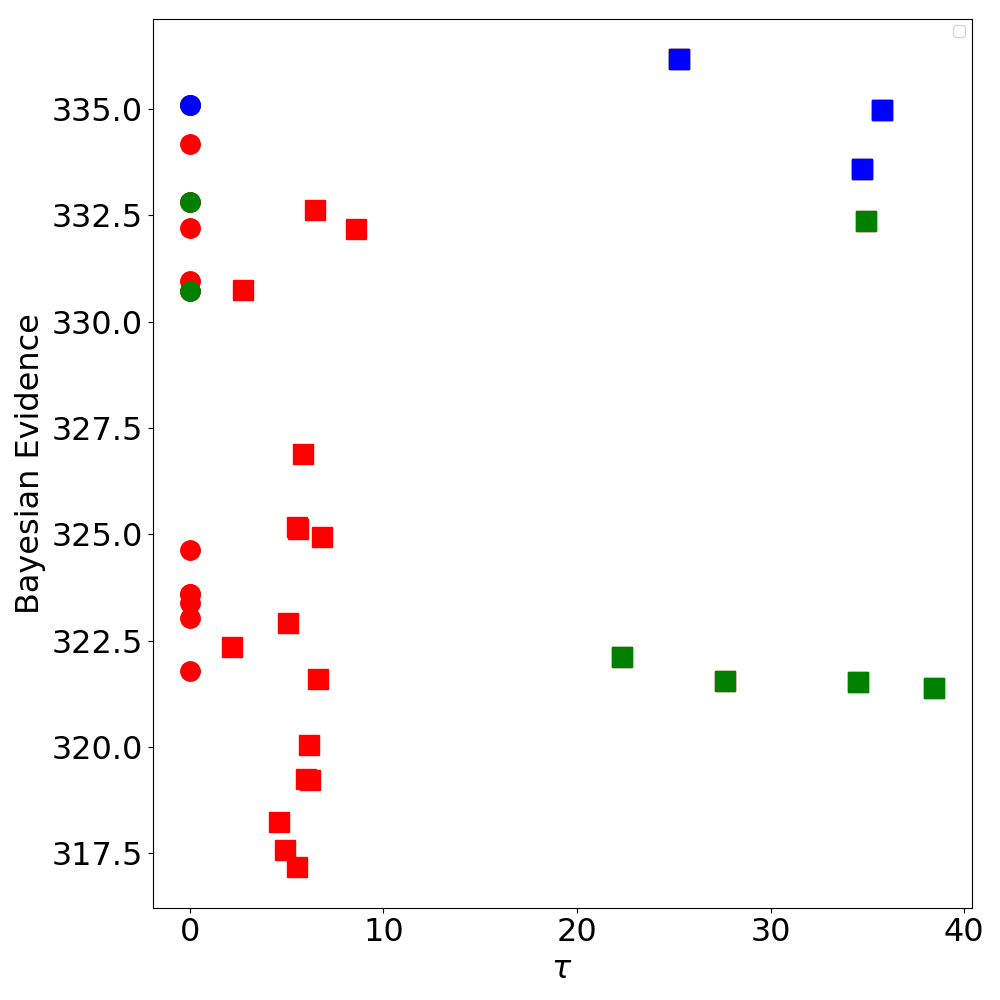}
	\includegraphics[width=0.5\textwidth]{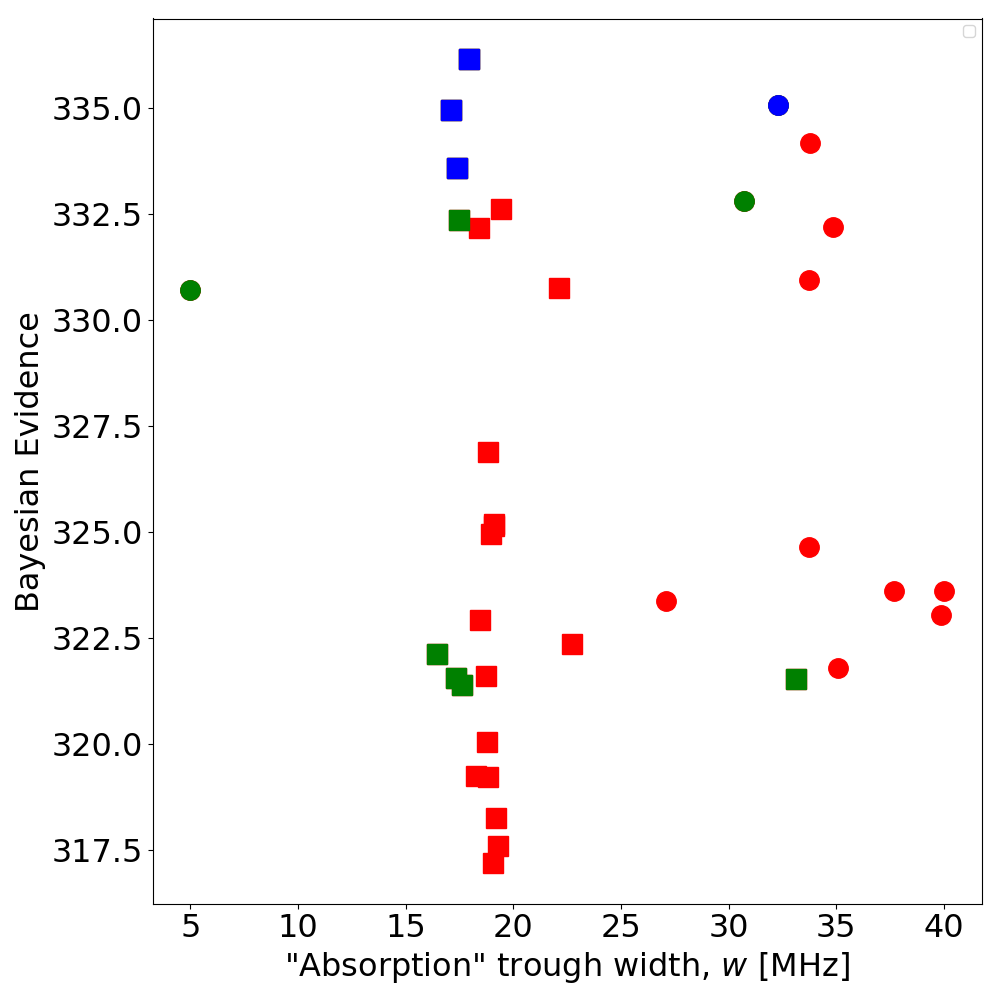}}
\caption{MAP values of the amplitude [tl], center frequency [tr], $\tau$ [bl] and width [br] of the absorption feature in models including a flattened Gaussian (squares) or Gaussian (circles) parametrisation of the global 21 cm signal vs. the Bayesian evidence of the model for models have Bayes factors of 20 or less relative to the best fitting model. Models with amplitudes that fit within the standard cosmological paradigm ($0<A<209~\mathrm{mK}$) and have Bayes factors of 3 or less, or greater than 3, relative to the best fitting model are shown in blue and green, respectively. Models with amplitudes that exceed that expected in the standard cosmological paradigm are shown in red. While, by definition, we do not explicitly fit for $\tau$ in models using a Gaussian parametrisation of the global 21 cm signal, in this figure we plot them as having a value of $\tau$ of zero, such that all of the models are represented in each subplot.}
\label{Fig:EvidencevsGSParams}
\end{figure*}

In \autoref{Fig:EvidencevsGSParams}, we plot the MAP values of the amplitude, centre frequency, $\tau$ and width of the absorption feature in models including either of the two Gaussian parametrisations of the global 21 cm signal considered in this paper, as a function of the Bayesian evidence of the model. For visual clarity, we limit models plotted to those that have Bayes factors of 20 or less relative to the best fitting model (models with log(evidence)>316). Models with amplitudes that exceed that expected in the standard cosmological paradigm ($A>209~\mathrm{mK}$) are shown in red. We split models with amplitudes that fit within the standard cosmological paradigm into two categories: those with a Bayes factors of 3 or less relative to m128 are shown in blue and lower evidence models are shown in green.
 
Of the five best fitting flattened Gaussian and Gaussian models, four have amplitudes consistent with those expected in the standard cosmological paradigm ($0<A<209~\mathrm{mK}$), with the one exception having a, physically less expected, extremely deep $4~\mathrm{K}$ absorption trough, which lies at the upper limit of the amplitude prior range. We also note that, of the non-Gaussian high evidence models, there are a further 4 {\sc{ARES}} models which have amplitudes consistent with those expected in the standard cosmological paradigm, by construction. Thus, in total, eight out of nine of the high evidence models that include global signal components have amplitudes in the range $0<A<209~\mathrm{mK}$ and, as such, marginalising over global 21 cm signal parametrisations, there is a clear preference for models with amplitudes consistent with those expected in the standard cosmological paradigm. 

Finally, we note that a cluster of lower evidence models, incorporating a flattened Gaussian parametrisation of the global 21 cm signal and having parameters similar to those derived in the analysis of B18 ($A\sim0.5~\mathrm{K}$, $\nu_{0}\sim78~\mathrm{MHz}$, $\tau \sim 7$ and  $A\sim18~\mathrm{MHz}$), are also clearly identifiable.

\section{Summary \& Conclusions}
\label{Conclusions}

Recently, B18 analysed data from the EDGES low-band radio antenna and found an unexpectedly deep absorption profile centred at 78 MHz, which may be a detection of the absorption signature at radio frequencies expected to accompany Cosmic Dawn. In their analysis, they fit their measurements using a polynomial foreground model, a flattened Gaussian absorption profile and a flat noise model.

In \autoref{EDGESBeamChromaticityCorrection}, we examined the beam chromaticity correction element of the EDGES low-band data calibration, and concluded that imperfections in the beam model or sky model used in that procedure have the potential to impart new (or leave residual) oscillatory spectral structure in the spectrum that could have biased estimates of the global signal in the data. Furthermore, we have argued that the flat noise model used by B18 (and also in other analyses of the publicly available EDGES low-band data by H18 and S19) is likely an oversimplification and that, instead, a generalised noise model, accounting for \begin{enumerate*}\item the expected radiometric noise on the data, \item the effect of RFI flagging on the data weights and \item additional noise-like structure that could plausibly be imparted by calibration errors or polarised foreground emission, is necessary to accurately describe the noise in the data.
\end{enumerate*}

We use a Bayesian evidence-based analysis of 128 models for the EDGES low-band data that include models for foreground emission and a global signal parametrised as a Gaussian, flattened Gaussian or {\sc{ARES}} simulation, as well models, of varying complexity (including no model at all), for systematic effects. Based on this analysis, we have shown that models incorporating components designed to describe a calibration systematic error dominated by an error in the amplitude, phase or period of the sinusoidal undulation component of the electromagnetic beam model for the gain at a particular beam angle and lower level contributions from a range of other beam angles, an imperfect beam chromaticity correction to the data, and potentially polarised foreground emission describe statistically important features of the data. We find that models incorporating components designed to describe these plausible calibration systematics and assuming a generalised noise model are decisively preferred by the Bayesian evidence (log-Bayes factors > 9 and 11 between best fitting models including both component and excluding either one, respectively) relative to those that do not.

Out of the total set of models analysed, the model with the highest Bayesian evidence (m128) includes a flattened Gaussian parametrisation for the global 21 cm signal, a 7th order log-polynomial model, power law damped sinusoid and generalised noise model. Additionally, there are a further twelve models (m116-m127) with Bayesian evidence values sufficiently high to not be strongly disfavoured relative to m128 ($\log(B)>3$). All of these models include higher order log-polynomial components than expected to be necessary to describe the foregrounds and ionosphere alone, a damped sinusoidal component and a noise level significantly in excess of that which is theoretically expected. This indicates the presence of spectral structure in the EDGES low-band data that is not consistent with exclusively including a combination of intrinsic foreground emission, or even foregrounds modulated by the ionosphere, an imperfectly calibrated-but-smooth-spectrum receiver gain and a global 21 cm signal.

Of the 13 highest evidence models, there are 3 with flattened Gaussian and 2 Gaussian global 21 cm signal parametrisations, 4 have no global signal component and 4 use global 21 cm signals derived from our atlas of 100,000 {\sc{ARES}} simulations. We thus conclude that neither the specific shape of the global 21 signal, nor, indeed, its detection, are preferred by a statistically significant margin in the publicly available EDGES low-band data set. Instead, the requirement for calibration systematic based models, in the form of an explicit power law damped sinusoidal feature, high-order (greater than 5th order) log-polynomials and a generalised noise model, is of greater statistical significance. While our inclusion of these components is motivated by the presence of calibration systematics, it is also possible that they are describing spectral structure in the data of another origin. However, what the model selection analysis does conclusively demonstrate is that models including components beyond those expected to be necessary to describe foregrounds, ionospheric effects and a global signal are decisively preferred.

Analyzing the global signal component of the data models in more detail, we find that, in the subset of the 13 highest evidence models models that include a global signal, 8 out of 9 have amplitudes that are consistent with standard cosmological expectations ($A<209~\mathrm{mK}$), in which the CMB is the dominant radio background at CD and the hydrogen gas cools adiabatically following decoupling from the CMB at $z\sim150$. This is in contrast to $A\sim500~\mathrm{mK}$ amplitude derived with the model used by B18. Focusing specfically on models that include Gaussian or flattened Gaussian parametrisations of the global signal, we find that the highest evidence models have $A<209~\mathrm{mK}$ and identify a cluster of lower evidence models incorporating a flattened Gaussian parametrisation of the global 21 cm signal, but lacking components designed to describe the effects of calibration systematics or  polarised foreground emission, that have global signal parameters similar to those derived in the analysis of B18 ($A\sim0.5~\mathrm{K}$, $\nu_{0}\sim78~\mathrm{MHz}$, $\tau \sim 7$ and  $A\sim18~\mathrm{MHz}$).

Ultimately, we conclude that a detection of  the global 21 cm signal is not strongly preferred by a statistically significant margin in the publicly available EDGES low-band data set, but if a signal is present, characterisation of its shape is limited by correlation with model components designed to describe calibration systematics in the data. However, the most strongly constrained feature of the signal is its amplitude, which has an evidence weighted MAP amplitude of $A<209~\mathrm{mK}$ and thus is entirely consistent with expectations of the standard cosmological model.

\section*{Acknowledgements}

PHS and JCP both acknowledge support from Brown University's Richard B. Salomon Faculty Research Award Fund. This research was conducted using computational resources and services at the Center for Computation and Visualization, Brown University. PHS thanks Irina Stefan for valuable discussions and helpful comments on a draft of this manuscript. We thank the EDGES collaboration for releasing the publicly available data set and J. Bowman and R. Monsalve for answering a number of questions related to the calibration of the EDGES data.



\label{lastpage}

\end{document}